\documentclass[fleqn,usenatbib]{mnras}

\usepackage{newtxtext,newtxmath}
\usepackage[T1]{fontenc}

\DeclareRobustCommand{\VAN}[3]{#2}
\let\VANthebibliography\thebibliography
\def\thebibliography{\DeclareRobustCommand{\VAN}[3]{##3}\VANthebibliography}

\usepackage{graphicx}	% Including figure files
\usepackage{amsmath}	% Advanced maths commands
\usepackage{caption}
\usepackage[flushleft]{threeparttable}
\title[Stratified PIC Simulations of the MRI]{Particle-in-cell Simulations of the Magnetorotational Instability \\ in Stratified Shearing Boxes}  

\author[A. Sandoval et al.]{Astor Sandoval,$^1$\thanks{E-mail: astor.sandoval@ug.uchile.cl} 
 Mario Riquelme,$^1$ Anatoly Spitkovsky$^2$ and Fabio Bacchini$^{3,4}$\\
$^{1}$Departamento de F\'isica, Facultad de Ciencias F\'isicas y Matem\'aticas, Universidad de Chile, Santiago, Chile\\
$^{2}$Department of Astrophysical Sciences, Princeton University, Princeton, NJ 08544, USA\\
$^{3}$Centre for mathematical Plasma Astrophysics, Department of Mathematics, KU Leuven, Celestijnenlaan 200B, B-3001 Leuven, Belgium\\
$^{4}$Royal Belgian Institute for Space Aeronomy, Solar-Terrestrial Centre of Excellence, Ringlaan 3, 1180 Uccle, Belgium}

\date{Accepted XXX. Received YYY; in original form ZZZ}

\pubyear{2023}

\begin{document}
\label{firstpage}
\pagerange{\pageref{firstpage}--\pageref{lastpage}}
\maketitle

\begin{abstract}
The magnetorotational instability (MRI) plays a crucial role in regulating the accretion efficiency in astrophysical accretion disks. In low-luminosity disks around black holes, such as Sgr A* and M87, Coulomb collisions are infrequent, making the MRI physics effectively collisionless. The collisionless MRI gives rise to kinetic plasma effects that can potentially affect its dynamic and thermodynamic properties. We present 2D and 3D particle-in-cell (PIC) plasma simulations of the collisionless MRI in stratified disks using shearing boxes with net vertical field. We use pair plasmas, with initial $\beta=100$ and concentrate on sub-relativistic plasma temperatures ($k_BT \lesssim mc^2$). Our 2D and 3D runs show disk expansion, particle and magnetic field outflows, and a dynamo-like process. They also produce magnetic pressure dominated disks with (Maxwell stress dominated) viscosity parameter $\alpha \sim 0.5-1$. By the end of the simulations, the dynamo-like magnetic field tends to dominate the magnetic energy and the viscosity in the disks. Our 2D and 3D runs produce fairly similar results, and are also consistent with previous 3D MHD simulations. Our simulations also show nonthermal particle acceleration, approximately characterized by power-law tails with temperature dependent spectral indices $-p$. For temperatures $k_BT \sim 0.05-0.3\, mc^2$, we find $p\approx 2.2-1.9$. The maximum accelerated particle energy depends on the scale separation between MHD and Larmor-scale plasma phenomena in a way consistent with previous PIC results of magnetic reconnection-driven acceleration. Our study constitutes a first step towards modeling from first principles potentially observable stratified MRI effects in low-luminosity accretion disks around black holes.
\end{abstract}

\begin{keywords}
plasmas -- instabilities -- accretion disks -- acceleration of particles -- dynamo
\end{keywords}

%%%%%%%%%%%%%%%%%%%%%%%%%%%%%%%%%%%%%%%%%%%%%%%%%%

%%%%%%%%%%%%%%%%% BODY OF PAPER %%%%%%%%%%%%%%%%%%

\section{Introduction}
\label{sec:intro}
\noindent The primary driver of accretion in astrophysical disks is believed to be the turbulence generated by the magnetorotational instability \citep[MRI;][]{BalbusEtAl1991,BalbusEtAl1998}, which provides the needed outward transport of angular momentum. Most of our knowledge about the nonlinear evolution of the MRI in different disk regimes comes from magnetohydrodynamic (MHD) simulations. However, in the regime where the plasma accretion rate is much lower than the Eddington rate, the Coulomb mean free path of the particles can be much larger than the system size, rendering the disk effectively collisionless and making the MHD approach inapplicable. This collisionless accretion regime is expected, for instance, in the low-hard state of X-ray binaries \citep{EsinEtAl1997} as well as around the central supermassive black holes of most nearby galaxies, including M87 and Sagittarius A* (Sgr A*) in our own Milky Way \citep{YuanEtAl2014}.\newline

\noindent The collisionless version of the MRI can give rise to several kinetic plasma phenomena, which may in turn affect its dynamics as well as the thermodynamic properties of the accreting plasma. These kinetic phenomena have been studied mainly via unstratified shearing-box MRI simulations, using either a fluid approach through kinetic-MHD models \citep{SharmaEtAl2006,SharmaEtAl2007} or particle simulations that employ either the hybrid or the particle-in-cell (PIC) methods \citep{RiquelmeEtAl2012,Hoshino2013,Hoshino2015,KunzEtAl2016,InchingoloEtAl2018,BacchiniEtAl2022}. One of the relevant kinetic phenomena is the appearance of an anisotropic stress, which is due to the presence of a pressure anisotropy in the accreting turbulence. Previous unstratified shearing-box simulation studies, both based on fluid and particle methods, have found that this anisotropic stress may contribute significantly to the disk viscosity, making the collisionless MRI turbulence more efficient in transporting angular momentum compared to its collisional counterpart. %However, kinetic-MHD results that consider a stratified disk (Hibakashi et al 2017) have found that disk stratification may decrease the contribution of anisotropic stress significantly. 
\newline

\noindent Another potentially important collisionless phenomenon is the possibly different ion and electron heating rates \citep[e.g.,][]{SharmaEtAl2007}. However, to date PIC studies have only used an ion to electron mass ratio $m_i/m_e=1$ (or close to unity), therefore not capturing the possibly different heating efficiencies of the different species. Plasma energization can also include nonthermal particle acceleration. Studying this phenomenon requires fully kinetic treatments of at least one species, which has been done through PIC and hybrid simulations. Different levels of nonthermal particle acceleration have indeed been found by these types of studies \citep{RiquelmeEtAl2012,Hoshino2013,Hoshino2015,KunzEtAl2016,InchingoloEtAl2018,BacchiniEtAl2022}, although the conditions under which this acceleration is most efficient and the mechanism(s) underlying this phenomenon remain to be clarified.\newline

\noindent An important physical ingredient, so far not included in hybrid or fully kinetic PIC studies of the MRI, is the vertical stratification of the disks. While the unstratified local shearing-box approximation allows us to investigate a disk by focusing on a small vertical section, this approach does not account for potentially important processes in stratified disks, such as outflows, disk expansion and the generation of a corona, among others. Stratified disks have been included in previous MHD shearing-box simulations of the MRI, which have found that stratification can give rise to important phenomena like outflows and dynamo-like processes, which may in turn affect the overall accretion efficiency of the disks \citep{BaiEtAl2013,SalvesenEtAl2016}. Also, a kinetic-MHD study that considers a stratified disk \citep{HirabayashiEtAl2017} has found that disk stratification may decrease the importance of anisotropic stress significantly compared to unstratified kinetic-MHD results. \newline

\noindent To address these possible effects, our study employs 2D and 3D stratified shearing-box PIC simulations to examine the development of the collisionless MRI. We use equal ion and electron masses, $m_i = m_e = m$ for computational convenience, and focus on the sub-relativistic temperature regime, relevant for the inner regions of black hole accretion disks ($k_B T \lesssim mc^2$, where $k_B$ is the Boltzmann constant, $T$ is the plasma temperature and $c$ is the speed of light). By comparing with unstratified PIC runs, we show the importance of including stratification to describe phenomena like plasma beta evolution, effective viscosity and particle acceleration. In our 2D runs we pay special attention to the role played by the ratio between the initial cyclotron frequency of the particles and the Keplerian frequency of the disk, $\omega_{c,0}/\Omega_0$ (hereafter, the scale-separation ratio). In realistic disks, this ratio satisfies $\omega_{c,0}/\Omega_0 \gg 1$ and determines the scale separation between mesoscale MHD phenomena and kinetic microphysical processes. Even though most of our analysis is done in 2D, in this paper we take a step forward by conducting the first fully kinetic 3D simulation of the stratified MRI evolution. This preliminary 3D simulation enables us to compare it with our established 2D results and gain valuable insight into the limitations of the 2D approach. This exploration sets the stage for future investigations aimed at fully unraveling the complexities of the 3D scenario.\newline %One important parameter in our simulations is the ratio between the initial cyclotron frequency of the ions and the Keplerian frequency of the disk, $\omega_{c,i}/\Omega_0$, which needs to be sufficiently large to ensure a proper separation between kinetic and MHD scales.\newline

\noindent The paper is organized as follows. In \S \ref{sec:setup} we describe our numerical method and simulation setup. In \S \ref{sec:2dmri} and \S \ref{sec:3dmri} we present the general properties of the stratified MRI turbulence in 2D and 3D, respectively. In \S \ref{sec:effectivevisc} we quantify the effective viscosity in our runs, and in \S \ref{sec:heataccel} we analyze the ability of the stratified MRI turbulence to accelerate particles. Finally, we present our conclusions in \S \ref{sec:conclu}.

\section{Simulation Setup}
\label{sec:setup}

We use the electromagnetic PIC code TRISTAN-MP \citep{Buneman1993, Spitkovsky2005} in 2D and 3D. Our simulations are performed in the local, shearing-box approximation \citep{HawleyEtAl1995}, using Cartesian coordinates where the $x$, $y$ and $z$ axes correspond to the radial, azimuthal (or toroidal) and vertical directions of the disk, respectively. This reference frame rotates with an angular velocity $\boldsymbol{\Omega}_0=\Omega_0 \hat{z}$, corresponding to the Keplerian angular velocity at a radius that coincides with the center of our simulation box. In order to model a stratified disk, we include the vertical component of the gravitational force produced by the central object, $-m\Omega_0^2 z \hat{z}$, and we initially set up an isothermal disk in hydrostatic equilibrium with a $z$-dependent density profile:\newline

\begin{equation}
	n(z)=n_0\exp\left(-\dfrac{z^2}{H_0^2}\right),
	\label{eq:hydrostat}	
\end{equation}
where $n_0$ is the plasma density at the disk midplane (considering both species), $H_0$ is the scale height of the disk given by $H_0=(2k_BT_0/m)^{1/2}/\Omega_0$ and $T_0$ is the initial plasma temperature, which is given by $k_BT_0/mc^2 = 5\times10^{-3}$ in all of our runs. Our runs do not include any type of particle cooling, so a gradual increase in the temperature and scale height of the simulated disks is expected due to dissipation of magnetic energy. The whole simulation domain is initially threaded by a vertical, homogeneous magnetic field $\boldsymbol{B}_0 = B_0\hat{z}$, so that the initial plasma $\beta$ parameter in the disk midplane, $\beta_0$ ($=8\pi n_0k_BT_0/B_0^2$), has a value of $\beta_0=100$. These choices for $T_0$ and $\beta_0$ imply that the initial Alfv\'en velocity in the disk midplane, $v_{A,0}$ ($=B_0/(4\pi n_0m)^{1/2}$), is $v_{A,0}/c=10^{-2}$ in all of our runs.

\subsection{Basic Equations}
\label{sec:equations}
\noindent In our rotating frame, the time derivative of particles momentum $\boldsymbol{p}=(p_x,p_y,p_z)$ is determined by the Lorentz force, the radial and vertical components of gravity, and the Coriolis force:\footnote{Since Coriolis forces conserve kinetic energy, the standard Coriolis expression for the evolution of $\boldsymbol{v}$, $d\boldsymbol{v}/dt = 2\boldsymbol{\Omega}_0 \times \boldsymbol{v}$, can be directly translated into a relativistic momentum $\boldsymbol{p}$ evolution as $d\boldsymbol{p}/dt = 2\boldsymbol{\Omega}_0 \times \boldsymbol{p}$.}
\begin{equation}
\frac{d\boldsymbol{p}}{dt} = q(\boldsymbol{E} + \frac{\boldsymbol{v}}{c}\times \boldsymbol{B}) + 3m\Omega_0^2 x \hat{x} -m\Omega_0^2 z \hat{z} - 2\boldsymbol{\Omega}_0 \times \boldsymbol{p} ,
\label{eq:force}
\end{equation}
where $\boldsymbol{v}=\boldsymbol{p}/(\gamma m)=(v_x,v_y,v_z)$, $q$, $\boldsymbol{E}$ and $\boldsymbol{B}$ are, respectively, the particle velocity, the particle charge and the electric and magnetic fields. In this non-inertial frame, Maxwell's equations also acquire extra terms, which modify the evolution of the electric field as \citep{Schiff1939}:   
\begin{equation}
\frac{\partial \boldsymbol{E}}{\partial t} = c\nabla \times \boldsymbol{B} -4\pi \boldsymbol{J} + \frac{\boldsymbol{v}_0}{c} \times \frac{\partial \boldsymbol{B}}{\partial t} - \nabla \times \Big(\boldsymbol{v}_0 \times \Big(\boldsymbol{E}-\frac{\boldsymbol{v}_0}{c}\times \boldsymbol{B}\Big)\Big),
\label{eq:mod_ampere}
\end{equation}
where $\boldsymbol{J}$ is the current density and $\boldsymbol{v}_0$ is the Keplerian rotation velocity of the disk at the center of our simulation box (the evolution of the magnetic field $\partial \boldsymbol{B}/\partial t  =  -c\nabla \times \boldsymbol{E}$ is not modified in the rotating frame). As discussed in \cite{RiquelmeEtAl2012}, the terms proportional to $\boldsymbol{v}_0$ in Eq.~\ref{eq:mod_ampere} can in principle be comparable to the displacement current $\partial \boldsymbol{E}/\partial t$, but should not change the non-relativistic MHD behavior of the plasma. This is because, in the non-relativistic regime ($|\boldsymbol{v}_0|=v_0 \ll c$), these extra terms are always much smaller than the first term on the right hand side of Eq.~\ref{eq:mod_ampere} ($c \nabla \times \boldsymbol{B}/4\pi$). Therefore, the current density $\boldsymbol{J}$ should still adjust to satisfy $\boldsymbol{J} \approx c \nabla \times \boldsymbol{B}/4\pi$, as assumed in the non-relativistic MHD approach. Thus, as it has been done in all previous PIC and hybrid studies of the MRI, we drop the terms proportional to $v_0$ in Eq.~\ref{eq:mod_ampere} and solve the conventional Maxwell's equations. We are thus implicitly assuming that these (beyond MHD) modifications to the displacement current do not affect considerably the kinetic MRI dynamics. 

\begin{table*}
\begin{center}
\begin{threeparttable}
%\tablecolumns{8}
%\tablewidth{0pt}

\caption{Simulations parameters \label{tab:param}}
%\tablehead{
\begin{tabular}{ccccccccc}
\hline \hline 
Run & UN2D-20& ST2D-28 & ST2D-20 & ST2D-14 & ST2D-10 & ST2D-7 &  ST2D-3.5 & ST3D-3.5\\ \vspace{-0.3cm}\\
\hline 
%\startdata
$\omega_{c,0}/\Omega_0$ & 20 & 28 & 20 & 14 & 10 & 7 & 3.5 & 3.5\\ 
$L_x$ [$2\pi v_{A,0}/\Omega_0$] & 22 & 35 & 43 & 47 & 46  & 46 & 48 & 24\\
$L_y$ [$2\pi v_{A,0}/\Omega_0$] & - & - & - & - & - & - & - & 24\\
$L_z$ [$2\pi v_{A,0}/\Omega_0$] & 22 & 120 & 89 & 95 & 92 & 93 & 96 & 96\\
$\Delta$ [$c/\omega_{p,0}$] & 0.35 & 0.35 & 0.35 & 0.35 & 0.35 & 0.35 & 0.35 & 0.35\\
$N_{ppc}$ & 25 & 400 & 350 & 200 & 200 & 200 & 200 & 30\\
$c$ [$\Delta/\Delta t$]& 0.45 & 0.45 & 0.45 & 0.45 & 0.45 & 0.45 & 0.45 & 0.225 \\
\hline
%$\omega_{c,0}/\Omega_0$ & 20 & 28 & 20 & 14 & 10 & 7 & 3.5 & 3.5\\ 
%$L_x$ [$2\pi v_{A,0}/\Omega_0$] & 22 & 22 & 22 & 22 & 45  & 45 & 45 & 16\\
%$L_y$ [$2\pi v_{A,0}/\Omega_0$] & - & - & - & - & - & - & - & 16\\
%$L_z$ [$2\pi v_{A,0}/\Omega_0$] & 22 & 22 & 22 & 22 & 90 & 90 & 90 & 90\\
%$\Delta$ [$c/\omega_{p,0}$] & 0.35 & 0.35 & 0.35 & 0.35 & 0.35 & 0.35 & 0.35 & 0.35\\
%$N_{ppc}$ & 50 & 50 & 25 & 25 & 200 & 200 & 300 & 200\\
%$c$ [$\Delta/\Delta t$]& 0.45 & 0.45 & 0.45 & 0.45 & 0.45 & 0.45 & 0.45 & 0.225 
\end{tabular}
\begin{tablenotes}
      \small
      \item  We list the initial parameters of our simulations, which are: the scale-separation ratio $\omega_{c,0}/\Omega_0$, where $\omega_{c,0} = |q|B_0/mc$ is the initial cyclotron frequency of the particles, the box size along the different axes ($L_x$, $L_y$ and $L_z$) in terms of $\lambda_{MRI}=2\pi v_{A,0}/\Omega_0$, the grid spacing $\Delta$ (equal in all dimensions) in terms of the initial plasma skin depth, $c/\omega_{p,0} = c/(4\pi n_0q^2/m)^{1/2}$, the initial number $N_{ppc}$ of particles (ions and electrons) per cell, and the speed of light $c$ in units of $\Delta/\Delta t$, where $\Delta t$ is the simulation time step.
      \end{tablenotes}
  \end{threeparttable}
  \end{center}
%\comments{We list the initial parameters of our simulations, which are: the scale-separation ratio $\omega_{c,0}/\Omega_0$, where $\omega_{c,0} = |q|B_0/mc$ is the initial cyclotron frequency of the particles, the box size along the different axes ($L_x$, $L_y$ and $L_z$) in terms of $\lambda_{MRI}=2\pi v_{A,0}/\Omega_0$, the grid spacing $\Delta$ (equal in all dimensions) in terms of the initial plasma skin depth, $c/\omega_{p,0} = c/(4\pi n_0q^2/m)^{1/2}$, the initial number $N_{ppc}$ of particles (ions and electrons) per cell, and the speed of light $c$ in units of $\Delta/\Delta t$, where $\Delta t$ is the simulation time step.}
\end{table*}

\subsection{Shearing Coordinates}
\label{sec:shearingcoordinates}
\noindent Simulating the MRI in the shearing-box approximation requires implementing shearing periodic boundary conditions in the radial ($x$) direction \citep[e.g.,][]{HawleyEtAl1995}. We do this by employing {\it shearing coordinates} \citep{RiquelmeEtAl2012}, in which the grid follows the shearing velocity profile within the shearing box, allowing the use of standard periodic boundary conditions in the radial ($x$) direction. However, the use of shearing coordinates introduces modifications in the evolution of the electric and magnetic fields, as well as in the evolution of particles momenta and positions. These modifications are described in detail in the Appendix of \cite{RiquelmeEtAl2012} and, for easy access, are also summarized below.\newline

\noindent In the shearing coordinates, the fields evolve as
\begin{eqnarray}
    \dfrac{\partial \boldsymbol{B}}{\partial t} & = & -c\nabla \times \boldsymbol{E} \label{eq:farad_comp}\\
    & &-\dfrac{3\Omega_0}{2}B_x\hat{y} + \dfrac{3\Omega_0}{2}\left( ct\dfrac{\partial \boldsymbol{E}}{\partial y} + \dfrac{y}{c} \dfrac{\partial \boldsymbol{E}}{\partial t} \right)\times \hat{x}\, \textrm{ and}\nonumber\\
    \dfrac{\partial \boldsymbol{E}}{\partial t}&=&c\nabla \times \boldsymbol{B}-4\pi\boldsymbol{J}  \label{eq:Amp_comp}\\ && -\dfrac{3\Omega_0}{2}E_x\hat{y}- \dfrac{3\Omega_0}{2}\left( ct\dfrac{\partial \boldsymbol{B}}{\partial y} + \dfrac{y}{c} \dfrac{\partial \boldsymbol{B}}{\partial t} \right)\times \hat{x}. \nonumber
\end{eqnarray}
%%%%%%%%%%%%%%%%% eq:farad_comp eq:Amp_comp
%the momentum \boldsymbol{p} evolves as:
%\begin{equation}
%\dfrac{d\boldsymbol{p}}{dt}=2\Omega_0p_y\hat{x}-\dfrac{1}{2}\Omega_0p_x\hat{y} - m\Omega_0^2z\hat{z} + q(\boldsymbol{E}+\dfrac{\boldsymbol{u}}{c}\times\boldsymbol{B}),
%\end{equation}
%whereas the evolution of the particles position $\boldsymbol{r} = (x,y,z)$ is given by:
%\begin{equation}
%\frac{d\boldsymbol{r}}{dt} = \boldsymbol{u} + \frac{3}{2}\Omega_0t\,u_x\hat{y}.
%\label{eq:dydt}
%\end{equation}
%The extra (time dependent) term in the evolution of the $y$ coordinate is explained by the time dependent shearing (or winding) of the grid as it follows the shearing motion of the plasma, which explains why a particle with $u_y=0$ and $u_x \ne 0$ should still change its $y-$coordinate. In Eq.~\ref{eq:dydt} we are assuming that the shear velocity $v$ in our shearing box is non-relativistic. This assumption is justified since $v \sim \Omega_0x$, and $x$ should be of the order of a few times the wavelength of the fastest growing MRI modes, $\lambda_{MRI} \approx 2\pi v_A/\Omega_0$ (where $v_A$ is the Alfven velocity of the plasma), which allows us to approximate $\Omega_0y \sim v_A$. This implies that Eq.~\ref{eq:dydt} is valid in the non-relativistic regime of the MRI turbulence.\newline
%%%%%%%%%%%%%%%%
The last terms in these equations, which are proportional to $y/c$ (hereafter, the {$y$-dependent terms}), can, however, be neglected in the $v_{A,0}/c \ll 1$ regime, as it is shown below. This can be seen considering that the size of our shearing-boxes in the $y$ direction should be typically a few times the wavelength of the most unstable MRI modes $\lambda_{MRI} = 2\pi v_{A,0}/\Omega_0$, which means that $\Omega_0y \sim v_{A,0}$ (the fact that $\lambda_{MRI}$ is the dominant scale of the MRI turbulence even in its nonlinear stage will be shown in \S \ref{sec:turbevol3d} and further discussed in \S \ref{sec:valid}). Also, assuming that the order of magnitude of the time derivative of any field component $f$ should satisfy $\partial f/\partial t \sim \Omega_0f$, one can calculate the ratios between the magnitudes of the $y-$dependent terms in Eqs.~\ref{eq:farad_comp} and \ref{eq:Amp_comp} and the left hand side of these equations, obtaining:
\begin{equation}
\frac{\Big| (\Omega_0y/c)\partial \boldsymbol{E}/\partial t\Big|}{\Big|\partial \boldsymbol{B}/\partial t\Big|} \sim \frac{|\boldsymbol{E}|}{|\boldsymbol{B}|} \frac{v_{A,0}}{c},
\label{eq:ratio_farad}
\end{equation}
for Eq.~\ref{eq:farad_comp} and
\begin{equation}
\frac{\Big| (\Omega_0y/c)\partial \boldsymbol{B}/\partial t\Big|}{\Big|\partial \boldsymbol{E}/\partial t\Big|} \sim \frac{|\boldsymbol{B}|}{|\boldsymbol{E}|} \frac{v_{A,0}}{c}
\label{eq:ratio_ampere}
\end{equation}
for Eq.~\ref{eq:Amp_comp}. Since in general $|\boldsymbol{E}|/|\boldsymbol{B}|\lesssim1$ (which is verified in \S \ref{sec:valid}), the right hand side of Eq.~\ref{eq:ratio_farad} is much smaller than unity as long as $v_{A,0}/c \ll 1$, implying that the $y$-dependent term in Eq.~\ref{eq:farad_comp} can be safely neglected. The right hand side of Eq.~\ref{eq:ratio_ampere}, on the other hand, is not necessarily $\ll 1$ since its value depends on the precise magnitude of the ratio $|\boldsymbol{E}|/|\boldsymbol{B}|$, which makes the $y-$dependent term in Eq.~\ref{eq:Amp_comp} not necessarily negligible. However, using the approximation $\nabla f \sim \lambda_{MRI}^{-1}f \sim (\Omega_0/v_{A,0})f$, we can calculate the ratio between the magnitude of this $y-$dependent term and an estimate of the magnitude of the first term on the right hand side of Eq.~\ref{eq:Amp_comp} ($c\nabla \times \boldsymbol{B}$), obtaining:
\begin{equation}
\frac{\Big| (\Omega_0y/c)\partial \boldsymbol{B}/\partial t\Big|}{\Big|c\nabla \times \boldsymbol{B}\Big|} \sim \Big(\frac{v_{A,0}}{c}\Big)^2.
\label{eq:ratio_ampere2}
\end{equation}
This implies that dropping the $y-$dependent term in Eq.~\ref{eq:Amp_comp} should not change the non-relativistic MHD behavior of the plasma, in which $\boldsymbol{J} \approx c \nabla \times \boldsymbol{B}/4\pi$, and is also consistent with our previous choice of ignoring the terms proportional to $v_0$ in Eq.~\ref{eq:mod_ampere}. Doing a similar analysis, we find that the ratio between the magnitudes of the third and the first terms on the right hand side of Eq.~\ref{eq:Amp_comp} is
\begin{equation}
\frac{ \Omega_0|E_x|}{\Big| c\nabla \times \boldsymbol{B}\Big|} \sim \frac{|\boldsymbol{E}|}{|\boldsymbol{B}|} \frac{v_{A,0}}{c},
\label{eq:consistency}
\end{equation}
 so, for consistency, we also neglect the former. Therefore, in our simulations we evolve the fields by solving the equations:
\begin{eqnarray}
\frac{\partial \boldsymbol{B}}{\partial t} & = & -c\nabla \times \boldsymbol{E} \label{eq:farad2}\\
&&-\dfrac{3}{2}\Omega_0B_x\hat{y} + \frac{3}{2}\Omega_0t\,c\frac{\partial \boldsymbol{E}}{\partial y}\times \hat{x} \, \textrm{ and }
 \nonumber \\
\frac{\partial \boldsymbol{E}}{\partial t} & = & c\nabla \times \boldsymbol{B}-4\pi\boldsymbol{J}  - \frac{3}{2}\Omega_0t\,c\frac{\partial \boldsymbol{B}}{\partial y}\times \hat{x}.
\label{eq:amp2}
\end{eqnarray}
  
\noindent In terms of particles evolution, in the shearing coordinates each particle's momentum $\boldsymbol{p}$ evolves as \citep{RiquelmeEtAl2012}:
\begin{equation}
\dfrac{d\boldsymbol{p}}{dt}=2\Omega_0p_y\hat{x}-\dfrac{1}{2}\Omega_0p_x\hat{y} - m\Omega_0^2z\hat{z} + q(\boldsymbol{E}+\dfrac{\boldsymbol{v}}{c}\times\boldsymbol{B}),
\label{eq:p}
\end{equation}
which is valid in the limit $\Omega_0y \sim v_{A,0} \ll c$ and as long as the shear velocity of the plasma within our simulation domain, $\boldsymbol{v_s}$, is non-relativistic. This last assumption is justified since $v_s \sim \Omega_0x$, and $x$ in our shearing-box is also of the order of a few times $\lambda_{MRI} = 2\pi v_{A,0}/\Omega_0$. This implies that $|\boldsymbol{v_s}| \sim \Omega_0x \sim v_{A,0}$, making Eq.~\ref{eq:p} valid in the regime $v_{A,0} \ll c$.\newline

\noindent Finally, the evolution of the particles position $\boldsymbol{r} = (x,y,z)$ is given by:
\begin{equation}
\frac{d\boldsymbol{r}}{dt} = \boldsymbol{v} + \frac{3}{2}\Omega_0t\,v_x\hat{y},
\label{eq:dydt}
\end{equation}
%The extra (time dependent) term in the evolution of the $y$ coordinate is explained by the time dependent shearing (or winding) of the grid as it follows the shearing motion of the plasma, which explains why a particle with $u_y=0$ and $u_x \ne 0$ should still change its $y-$coordinate. 
which is obtained combining Eqs. A30 and A35 of \cite{RiquelmeEtAl2012} also in the limit in which $\boldsymbol{v_s}$ is non-relativistic.\newline 

\noindent In order to safeguard the numerical stability and accuracy of our simulations, every time the factor $(3/2)\Omega_0t$ on the right hand side of Eqs.~\ref{eq:farad2}, \ref{eq:amp2} and \ref{eq:dydt} equals an integer, we reset these equations to their initial ($t=0$) shape. This implies a periodic ``unshearing" of our shearing grid that, therefore, requires a remapping of the electric and magnetic fields, as well as of the particles positions. This periodic redefinition of the time origin in our runs means that the factors $(3/2)\Omega_0t$ in Eqs.~\ref{eq:farad2}, \ref{eq:amp2} and \ref{eq:dydt} never surpass unity. Notice also that Eq.~\ref{eq:dydt} implies that relativistic particles may in principle change position in the $y-$direction at a rate close to twice the speed of light. This should not be considered a violation of special relativity, since this equation only describes the update of particle positions in our non-inertial, time-varying shearing coordinates. However, this situation may affect the numerical stability of our method. In order to avoid this possibility, our 3D runs use $c=0.225 \Delta/\Delta t$. Since this is not an issue in our 2D runs, in those cases we use $c=0.45 \Delta/\Delta t$ (see Table \ref{tab:param}). \newline
 
\noindent We emphasize that, in order to obtain our plasma evolution equations, we have assumed a non-relativistic plasma with $v_{A,0}/c \ll 1$, which rotates at non-relativistic velocities ($v_0 \ll c$). This implies that our work strictly applies to a plasma at radii significantly larger than the gravitational radius of a central black hole. For this reason, in this work we concentrate on a sub-relativistic regime, where the plasma temperature satisfies $k_BT \lesssim mc^2$. Notice, however, that the treatment of individual particles is relativistic, since (as we see below) a small fraction of them can still be nonthermally accelerated to energies much larger than $mc^2$.\newline

\subsection{Boundary conditions along $z$}
\label{sec:boundary}

\noindent Using shearing coordinates allows the use of periodic boundary conditions both in the $x$ (radial) and $y$ (toroidal) coordinates. In the $z$ coordinate we use open boundary conditions, which allow the existence of field and particle outflows in our stratified setup. Thus, in our runs particles are removed from the simulation box after they cross the vertical boundaries, while the fields are absorbed by these boundaries. This configuration effectively prevents outflowing fields from rebounding into the simulation domain \citep{CeruttiEtAl2015,Belyaev2015,SironiEtAl2016}. This is done by implementing an absorption layer of width $\Delta_{abs}$ in the vertical boundaries, where the terms
\begin{equation}
-\eta(z)(\boldsymbol{B}-\boldsymbol{B}_0)\,\,\,\,\textrm{ and } -\eta(z)\boldsymbol{E}
\end{equation}
are added to the right hand side of Eqs.~\ref{eq:farad2} and \ref{eq:amp2}, respectively. We use $\Delta_{abs} =50$ cells and $\eta(z)=(40/\Delta t)(|z-z_{abs}|/\Delta_{abs})^3$ within the absorption layer ($\eta(z)=0$ otherwise), where $z_{abs}$ is the inner edge of the absorption layer and $\Delta t$ is the simulation time step.

\subsection{Numerical parameters}
\label{sec:numerical}
\noindent The simulations presented in this paper and their numerical parameters are listed in Table \ref{tab:param}, with all physical quantities in stratified runs corresponding to plasma conditions in the disk midplane. These parameters are the scale-separation ratio $\omega_{c,0}/\Omega_0$, where $\omega_{c,0} = |q|B_0/mc$ is the initial cyclotron frequency of the particles, the box size along the different axes ($L_x$, $L_y$ and $L_z$) in terms of $\lambda_{MRI}$, the grid spacing $\Delta$ in terms of the initial plasma skin depth, $c/\omega_{p,0} = c/(4\pi n_0q^2/m)^{1/2}$, the initial number $N_{ppc}$ of macro-particles (ions and electrons) per cell and the speed of light $c$ in units of $\Delta/\Delta t$, where $\Delta t$ is the simulation time step. Notice that we ran simulations using several values of $\Delta$, $N_{ppc}$ and $L_x$, $L_y$ and $L_z$ to make sure that our results are numerically converged. Table \ref{tab:param} only includes the simulations used to present our results. 
%MRI. Table \ref{tab:param} shows the parameter of the simulations, where the H simulation correspond to our fiducial unstratified case. While SA, SB and SC correspond to the stratificated scenario where the conditions in the middle of the disk are the same as the unstratified scenario. \cite{2018ApJ...859..149I} shows that high spatial domain (solve various MRI modes) is mandaroty to reach numerical convergence. However, in a real disk the number of MRI modes is determined by the initial $\beta$ parameter. 
%\subsection{Unstratified shearing-box} \label{subsec:uns}
\subsection{Notation Convention}
\label{sec:notation}

\noindent In this section, we introduce various types of averages denoted by angled brackets with different subscripts, namely $\langle A\rangle_x$, $\langle A\rangle_{x-y}$, and $\langle A\rangle_{v}$. $\langle A\rangle_x$ denotes the average along the $x$ axis at a fixed height $z$ for 2D stratified simulations, $\langle A\rangle_{x-y}$ denotes the average over the $x-y$ plane at a fixed $z$ for 3D stratified simulations, and $\langle A\rangle_v$ represents the average taken over the volume of the disk for stratified simulations, while for unstratified simulations it represents the average over the entire simulation domain. \newline %of the quantity $A(x,y,z)$, which are given by:
%\begin{equation}
%\langle A\rangle_x=\dfrac{1}{L_x}\int_0^{L_x}Adx,
%\end{equation}
%\begin{equation}
%\langle A\rangle_{x-y}=\dfrac{1}{L_xL_y}\int_0^{L_x}\int_0^{L_y}Adxdy,
%\end{equation}
%\begin{eqnarray}
%\langle A\rangle_v=\left\{ \begin{array}{ll}
%			  \dfrac{\int_{0}^{L_z}\int_0^{Lx} A dx dz}{L_xL_z}, &  \text{ for 2D Unstratified   }\vspace{-2mm} \\	
%			 &\text{ simulations}\\
%             \dfrac{\int_{-H}^{H}\int_0^{Lx} A dx dz}{2L_xH}, &  \text{ for 2D Stratified   } \vspace{-2mm}\\
%            & \text{ simulations} \\
%             \dfrac{\int_{-H}^{H}\int_0^{Ly}\int_0^{Lx}Adxdydz}{2L_xL_yH}, & \text{ for 3D stratified   }\vspace{-2mm}\\
%             & \text{ simulations}
%             \end{array}
%   \right.
%\end{eqnarray}
%$\overline{\sigma_c}=\langle B^2\rangle_v/\langle 4\pi n m c^2\rangle_v$
%$\overline{\alpha}=\langle T_{xy}\rangle_v/\langle P \rangle_v$

\noindent Additionally, we use an overline notation ($\overline{\phantom{x}}$) for quantities that are computed as the ratio of two volume averages. For instance, for the plasma $\beta$ and temperature we define $\overline{\beta}\equiv\langle 8\pi P \rangle_v/\langle B^2 \rangle_v$ and $k_B\overline{T}\equiv\langle P \rangle_v/\langle n\rangle_v$. In these expressions, $\langle P\rangle_v=\langle P_{\parallel}\rangle_v/3+2\langle P_{\perp}\rangle_v/3$, where $P$ denotes the isotropic pressure, and $P_{\parallel}$ and $P_{\perp}$ correspond to the pressure parallel and perpendicular to the local magnetic field, respectively. \newline

\noindent Since in the stratified runs these averages are calculated in the disk region, we define this region through the condition $|z| < H(\overline{ T})$, where $H(\overline{ T})$ $=(2k_B \overline{ T}/m)^{1/2}/\Omega_0$ denotes the instantaneous scale height of the disk. Notice that the calculation of $\overline{T}$ has to be done in the disk region itself, whose definition depends on $\overline{ T}$ through the inequality $|z| < H(\overline{ T})$, implying that the $\overline{ T}$ of the disk has to be determined recursively.
\begin{figure} 
%\vspace{0.5cm} 
\hspace*{0cm} \includegraphics[width=1\columnwidth]{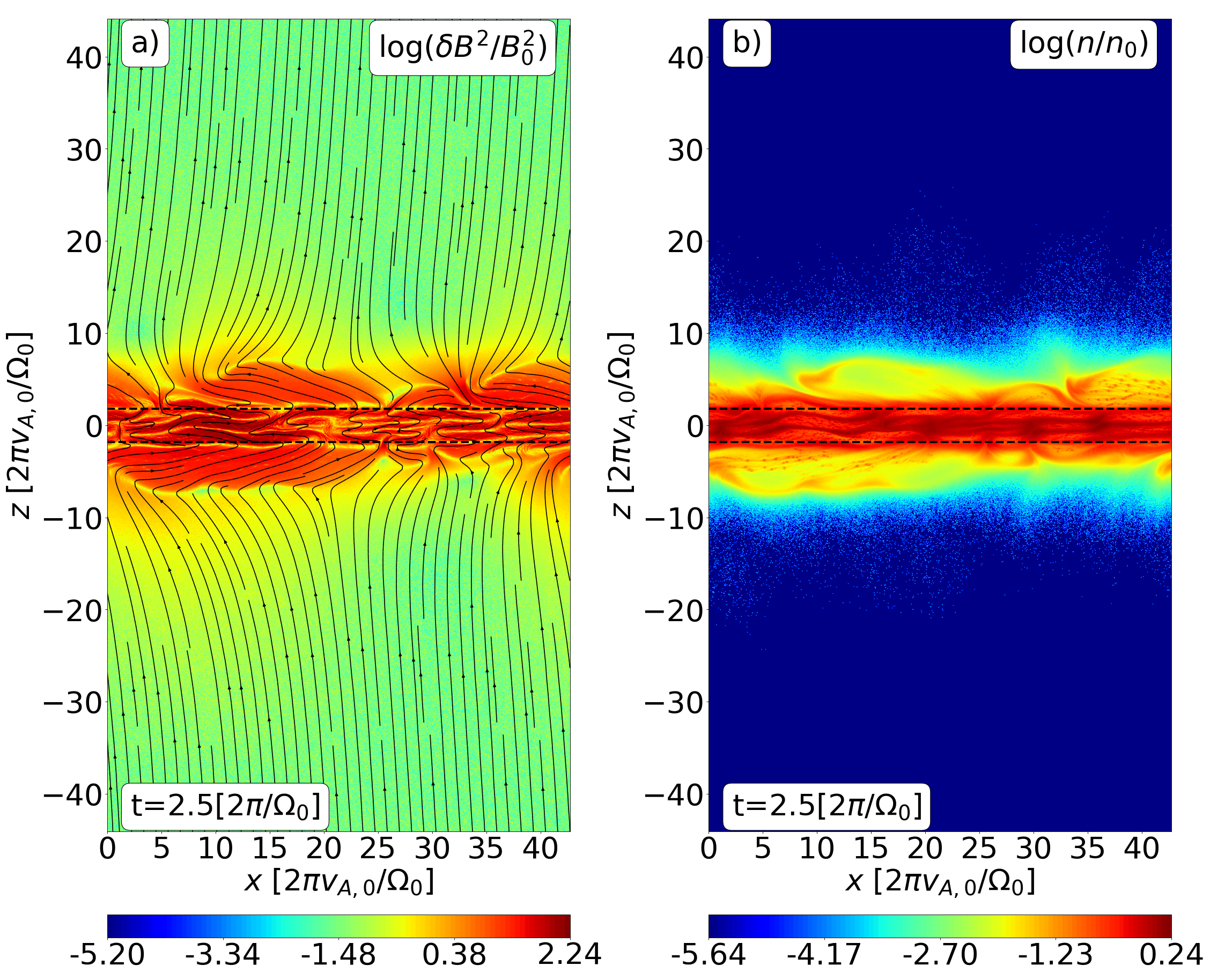}\\
\hspace*{0cm} \includegraphics[width=1\columnwidth]{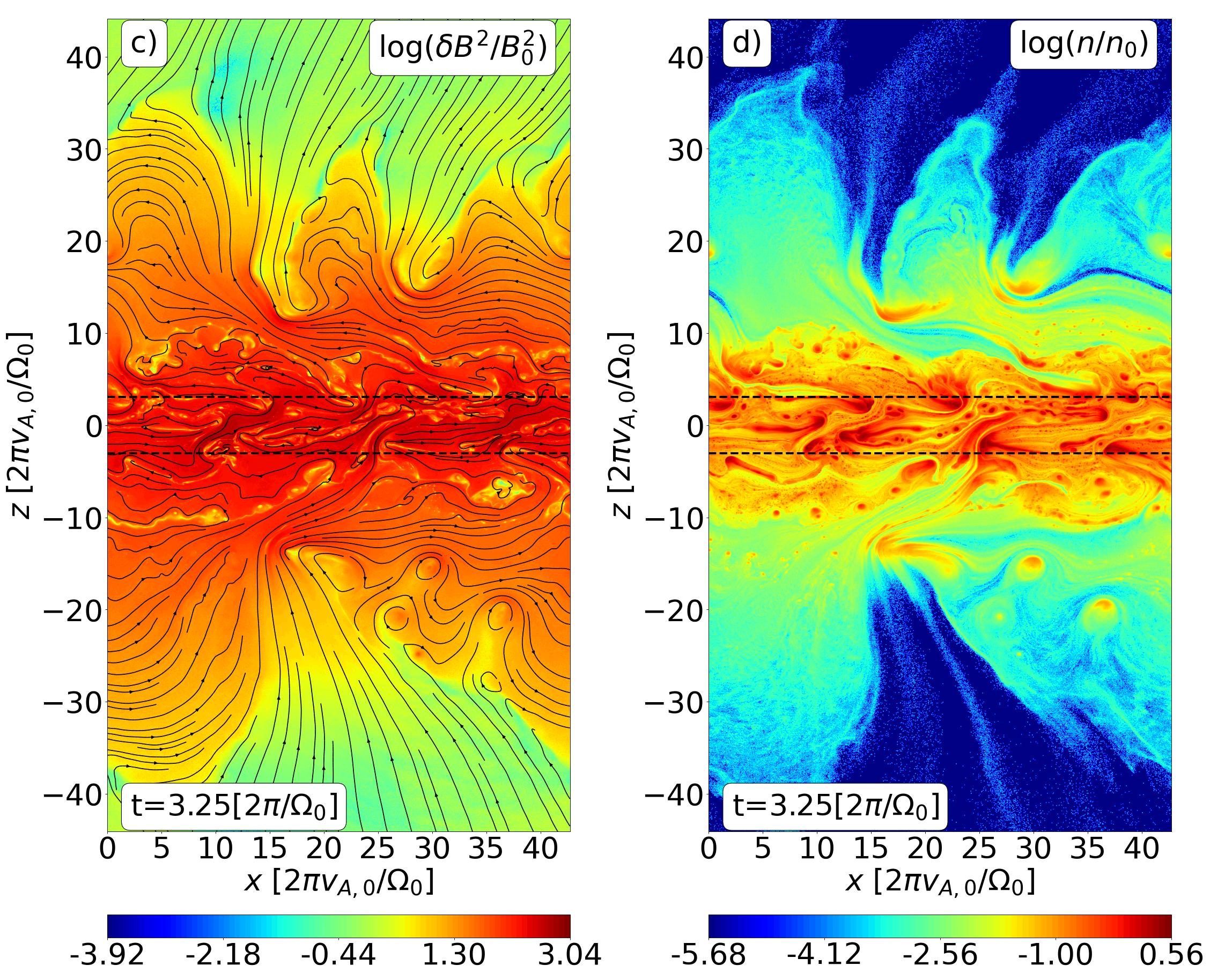}\\
\hspace*{0cm} \includegraphics[width=1\columnwidth]{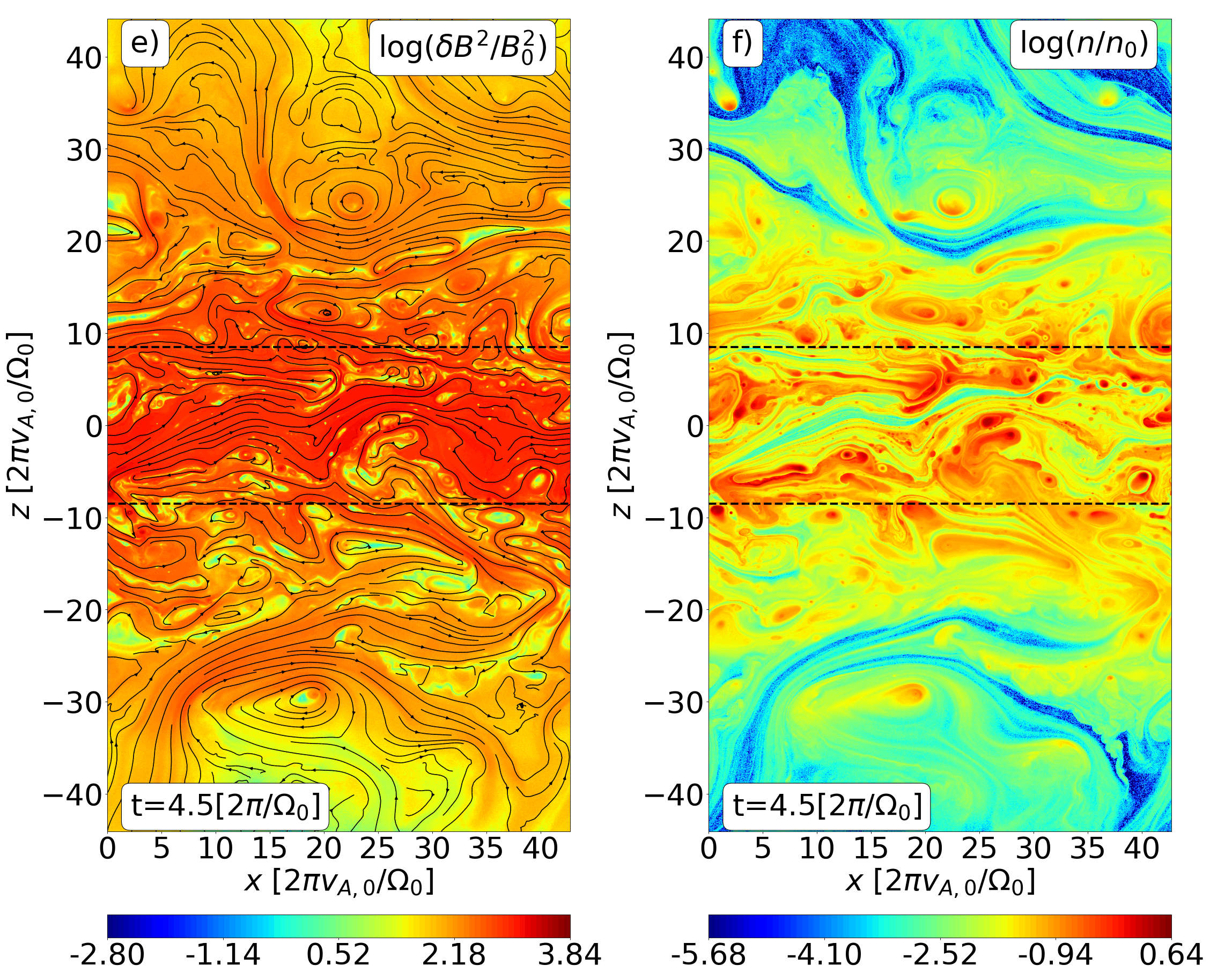}
\caption{Squared magnetic fluctuations $\delta B^2$ (left) and plasma density $n$ (right) for simulation ST2D-20 at $t=2.5$, 3.25 and 4.5 [$2\pi/\Omega_0$]. The black arrows in the left panels show the total magnetic field direction. The horizontal dotted lines in all the panels mark the region defined as disk region in our analysis (i.e., $|z| < H(\overline{ T})$).}
    \label{fig:strat_snaps}
\end{figure}
\begin{figure}
    \centering
%\vspace*{0.1cm}
\hspace*{-0.cm} \includegraphics[width=.95\columnwidth]{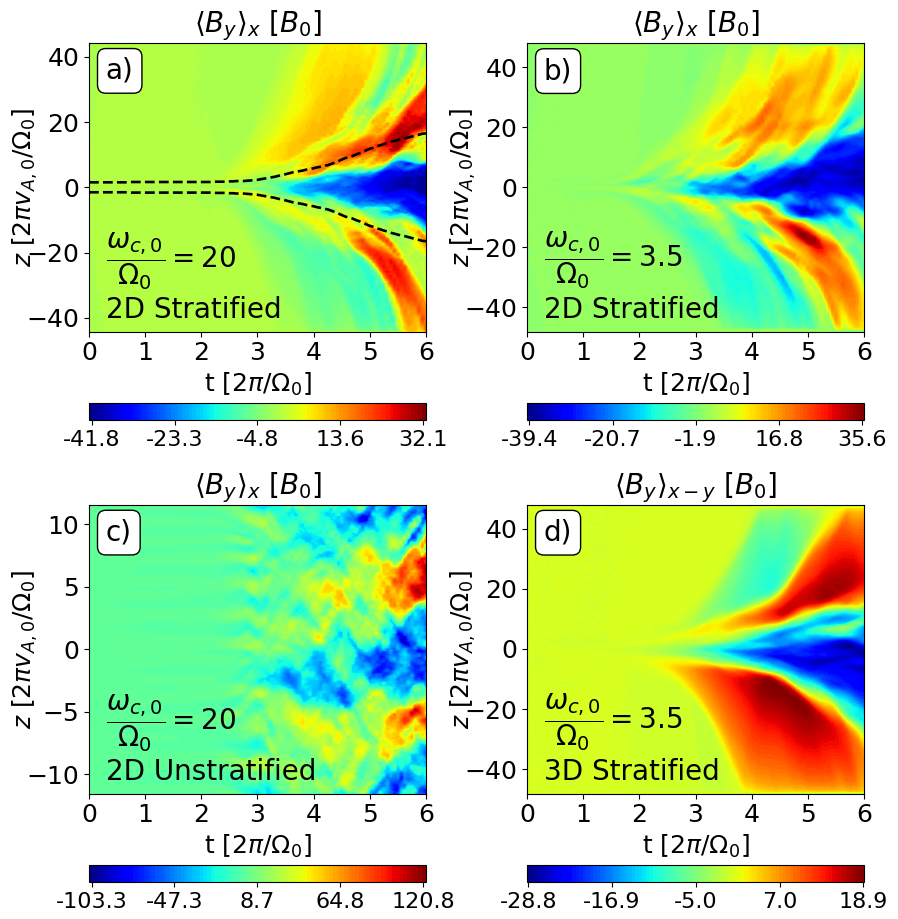}
%\vspace*{-1cm}
%    \includegraphics[width=.45\columnwidth]{fig13.png}
\caption{Panels $a$, $b$ and $c$ show the toroidal magnetic component $B_y$ averaged over the $x$-axis, $\langle B_y \rangle_x$, as a function of the time $t$ and the vertical coordinate $z$ for the 2D runs ST2D-20, ST2D-3.5 and UN2D-20, respectively. The black dashed line in panel $a$ marks the disk region ($|z| < H(\overline{ T})$). Similarly, panel $d$ shows $B_y$ averaged over the $x-y$ plane for the 3D run ST3D-3.5.}     \label{fig:dynamo2d}
\end{figure}

\section{2D results}
\label{sec:2dmri}

\noindent In this section we describe the stratified MRI turbulence using 2D simulations, paying special attention to the difference between stratified and unstratified simulations and to the role played by the scale-separation ratio $\omega_{c,0}/\Omega_0$. In \S \ref{sec:turbevol} we analyze the properties of the turbulence, and in \S \ref{sec:volume} we show the evolution of the plasma properties in the disk.

\subsection{Turbulence properties in 2D}
\label{sec:turbevol}
\noindent Figure \ref{fig:strat_snaps} shows three snapshots of the squared magnetic fluctuations $\delta B^2$ (where $\delta B = |\delta \boldsymbol{B}|$ and $\delta \boldsymbol{B}=\boldsymbol{B}-\boldsymbol{B}_0$) and of the particle density $n$ for the stratified 2D run ST2D-20 ($\omega_{c,0}/\Omega_0=20$). Panels $a$ and $b$ show the initial formation of nonlinear channel flows at time $t=2.5$ $[2\pi/\Omega_0]$. These channel flows appear both in $\delta B^2$ and $n$ and are more clearly formed within the disk region ($|z| < H(\overline{ T})$), which is marked by the horizontal dotted lines in all the panels. Panels $c$ and $d$ show the same quantities but at $t=3.25$ $[2\pi/\Omega_0]$, when the channel flows have already experienced reconnection, breaking into a turbulent state. At that moment, the disk thickness has increased due to plasma heating and significant particle and magnetic field outflows occur. This turbulent state continues during the entire simulation and is accompanied by a permanent puffing up of the disk, as shown by panels $e$ and $f$, corresponding to $t=4.5$ $[2\pi/\Omega_0]$. \newline

\noindent Our 2D runs also show the formation of a large scale, preferentially toroidal dynamo-like field, similar to those observed in previous MHD studies \citep[e.g.,][]{BaiEtAl2013,SalvesenEtAl2016}. This is seen in panel $a$  of Fig.~\ref{fig:dynamo2d}, which shows $\langle B_y \rangle_x$ as a function of time $t$ and of the vertical coordinate $z$. We see that a net $\langle B_y \rangle_x$ is formed, with a maximum amplitude of $\sim 30-40$ $B_0$ during the nonlinear stage of the stratified simulation and with oposite signs inside and outside the disk. The amplitude of $\langle B_y \rangle_x$ 
%In contrast, as depicted in panel $d$, the unstratified simulation also generates a mean $\langle B_y \rangle_x$ field, exhibiting a maximum amplitude of  $\sim 120$ $[B_0]$, accompanied by a sign variation over a vertical scale of $z \approx 5$ $[2\pi v_{A,0}/\Omega_0]$. Remarkably, the amplitude of $\langle B_y \rangle_x$ within our stratified simulation 
is very close to the one observed by previous equivalent 3D MHD simulations of the stratified MRI with initial $\beta_0=100$ in the disk midplane \citep{SalvesenEtAl2016}. \newline %In \S \ref{sec:volume} we show that this dynamo-like component of the magnetic field is indeed the dominant magnetic component in the nonlinear regime of the MRI turbulence.\newline 
\begin{figure} 
%\vspace{0cm} 
\hspace*{0cm} \includegraphics[width=1\columnwidth]{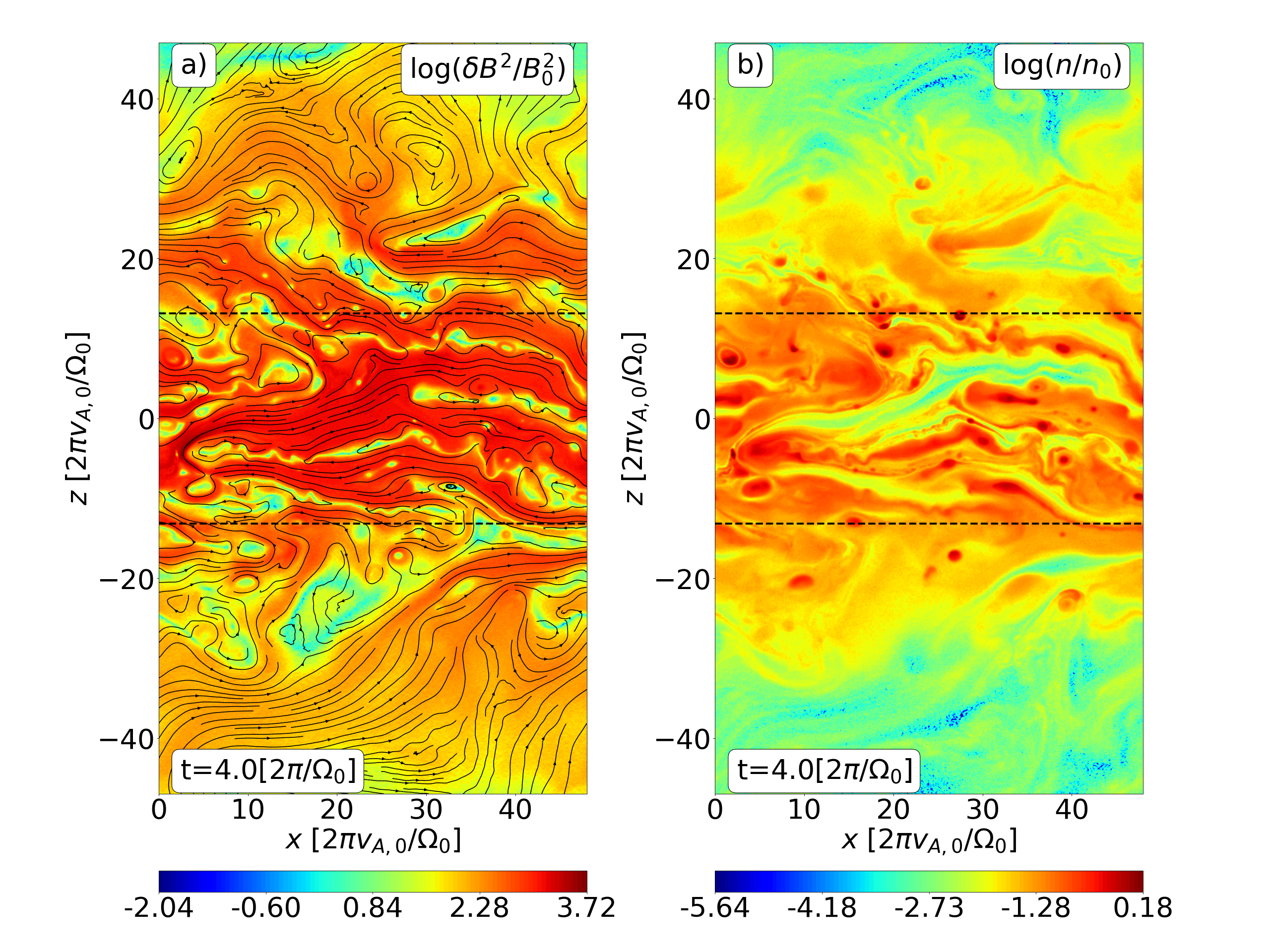}
%\vspace{-0.5cm} 
\caption{Panels $a$ and $b$ are analogous to panels $e$ and $f$ of Fig.~\ref{fig:strat_snaps} but for a run using a much smaller scale-separation ratio $\omega_{c,0}/\Omega_0=3.5$ (run ST2D-3.5). }
    \label{fig:strat_snap_3.5}
\end{figure}
%\noindent The presence of outflows, disk expansion and of a dynamo-like process has several effects on the properties of the disk. In order to assess these effects, we compared run ST2D-20 with an analogous unstratified simulation, run UN2D-20, where the initial conditions are the same as in the disk midplane of run ST2D-20. 
%\noindent Figure \ref{fig:uns_snaps} shows $\delta B^2$ and $n$ for run UN2D-20 at the moment when nonlinear channel flows appear ($t=2.5$ $[2\pi/\Omega_0]$) and then when these channel flows have reconnected and broken into turbulence ($t=3$ $[2\pi/\Omega_0]$). At first glance, this figure suggests that the evolution of the MRI turbulence in run UN2D-20 is similar to the one in the disk region of run ST2D-20. However, a more quantitative analysis of the disk plasma properties (\S \ref{sec:volume}) shows that the turbulence present in the unstratified run UN2D-20 has significant differences compared to the disk region of the stratified simulation ST2D-20.\newline

\noindent In order to explore the effect of the scale-separation ratio $\omega_{c,0}/\Omega_0$ on the 2D turbulence structure, Fig.~\ref{fig:strat_snap_3.5} shows $\delta B^2$ and $n$ in the nonlinear MRI state ($t=4$ [$2\pi/\Omega_0$]) for an analogous run using $\omega_{c,0}/\Omega_0=3.5$ (run ST2D-3.5) instead of $\omega_{c,0}/\Omega_0=20$. By comparing with panels $e$ and $f$ of Fig.  \ref{fig:strat_snaps}, we see that the scale-separation ratio does not appear to produce a qualitative change in the properties of the 2D MRI turbulence, preserving features such as disk thickness increase and the presence of outflows. Run ST2D-3.5 also shows significant dynamo-like activity, as seen in panel $b$ of Fig.~\ref{fig:dynamo2d}, where a net $\langle B_y \rangle_x$ field appears similarly to the case of run ST2D-20 in panel $a$.\newline

\begin{figure}
    \centering
    \hspace*{-0.5cm} \includegraphics[width=1\columnwidth]{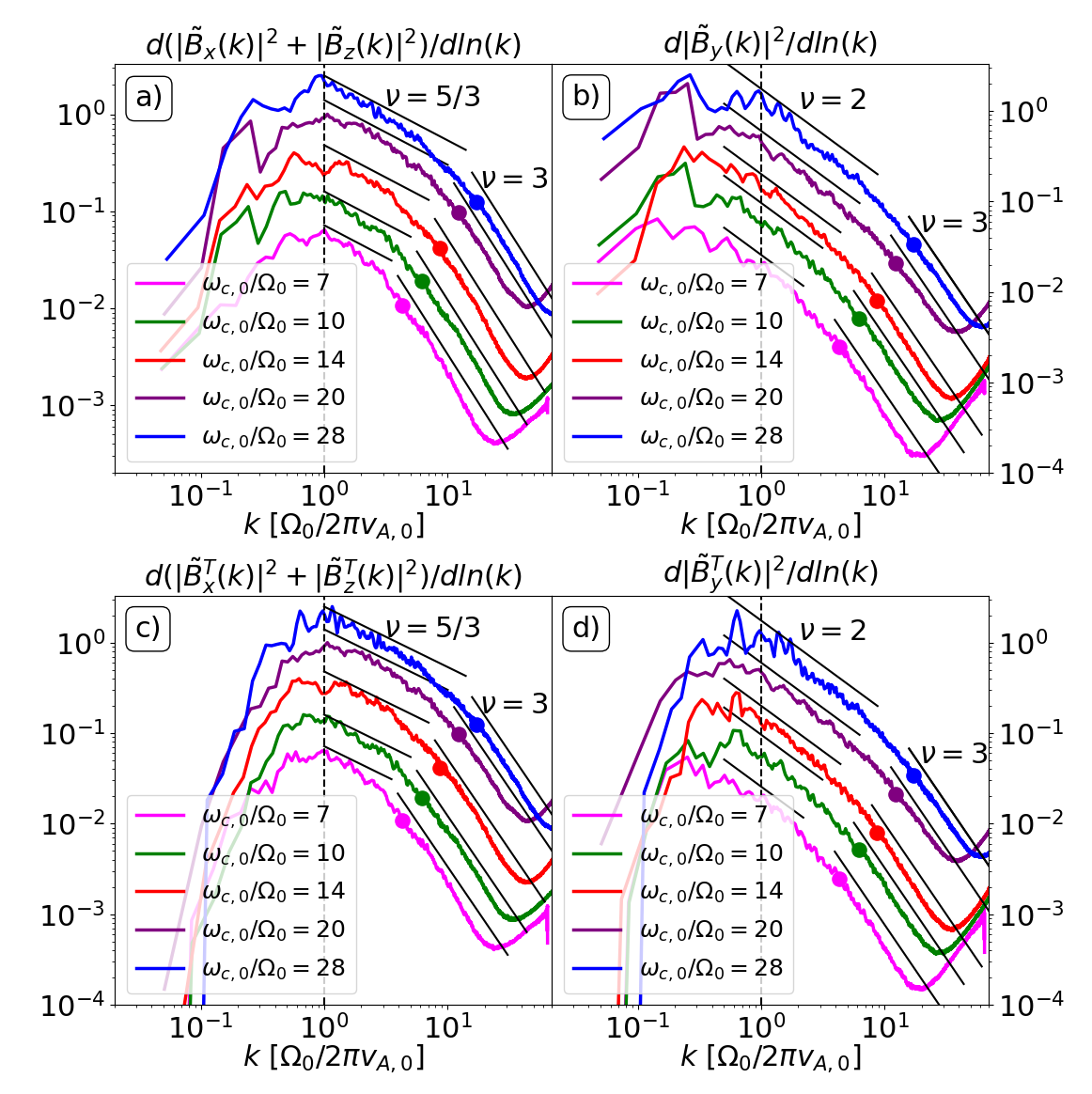}
%    \vspace{-0.5cm}
    \caption{Panels $a$ and $b$ show the power spectra of the poloidal and toroidal components of the total magnetic field, $d(|\tilde{B}_x(k)|^2+|\tilde{B}_z(k)|^2)/dln(k)$ and $d|\tilde{B}_y(k)|^2/dln(k)$, for 2D stratified runs with $\omega_{c,0}/\Omega_0$=7 (pink), 10 (green), 14 (red), 20 (purple) and 28 (blue). Panels $c$ and $d$ are analogous to panels $a$ and $b$, but considering only the turbulent component of the magnetic field, $\mathbf{B}^T$. In all the panels, the spectra use arbitrary normalization and the wave number $k$ is normalized by  $\Omega_0/2\pi v_{A,0}$.}
\label{fig:powerspectrum}
\end{figure}

\noindent The weak effect of $\omega_{c,0}/\Omega_0$ on the field structure of the stratified MRI can also be seen in the magnetic field power spectra, which are shown in Fig.~\ref{fig:powerspectrum} for runs with $\omega_{c,0}/\Omega_0=7, 10, 14$, 20 and 28 (all of them at $t\approx 5$ [$2\pi/\Omega_0$]). Panel $a$ shows the spectra of the poloidal component of the magnetic field, $d(|\tilde{B}_x(k)|^2+|\tilde{B}_z(k)|^2)/dln(k)$ ($\tilde{B}_x(k)$ and $\tilde{B}_z(k)$ are the Fourier transform of the $x$ and $z$ components of $\boldsymbol{B}$ and $k$ is the corresponding wave number), while panel $b$ shows the spectra of the toroidal component, $d|\tilde{B}_y(k)|^2/dln(k)$. For all the values of $\omega_{c,0}/\Omega_0$, the spectra show similar shapes, with a break at $k\rho_l \sim 1$ ($k\rho_l = 1$ is marked by the colored dots on each line), where $\rho_l$ is the typical particle Larmor radius, defined as $\rho_l\equiv mc(3k_B \overline{T}/m)^{1/2}/|q|\langle B^2 \rangle_v^{1/2}$. Their main difference is the location of the break of the spectra, which moves to larger wave numbers (in units of $\lambda_{MRI}^{-1}=\Omega_0/2\pi v_{A,0}$) as $\omega_{c,0}/\Omega_0$ increases, implying a growing separation between the kinetic ($\rho_l $) and the MHD ($\lambda_{MRI}$) scales. However, apart from this growing separation between scales, increasing $\omega_{c,0}/\Omega_0$ does not significantly affect the qualitative shape of the power spectra.\newline
\noindent Panels $a$ and $b$ also compare $d(|\tilde{B}_x(k)|^2+|\tilde{B}_z(k)|^2)/dk$ and $d|\tilde{B}_y(k)|^2/dk$ with power-law functions of index $\nu$ ($\propto k^{-\nu}$) and show that, at sub-Larmor scales ($k\rho_l \gtrsim 1$), the poloidal and toroidal spectra are approximately consistent with $\nu \approx 3$. This $\nu \approx 3$ behavior is expected for kinetic Alfv\'en wave turbulence \citep[e.g.,][]{PassotEtAl2015} and it has also been observed in previous unstratified 2D and 3D kinetic simulations \citep{KunzEtAl2016,InchingoloEtAl2018,BacchiniEtAl2022}. Above Larmor scales ($k\rho_l < 1$) the poloidal spectra show a peak at $k\,2\pi v_{A,0}/\Omega_0\sim 1$, followed by a power-law region characterized by $\nu\approx 5/3$. The toroidal spectra, on the other hand, has a peak at $k\,2\pi v_{A,0}/\Omega_0\sim 0.2$, followed first by an approximately flat region for $0.2 \lesssim k\,2\pi v_{A,0}/\Omega_0\lesssim 1$ and then by a steeper $\nu\approx 2$ region for $k\,2\pi v_{A,0}/\Omega_0 \gtrsim 1$. The nearly flat behavior of the toroidal spectra at $0.2 \lesssim k\,2\pi v_{A,0}/\Omega_0\lesssim 1$ is significantly affected by the presence of the dynamo-like field. Indeed, panels $c$ and $d$ of Fig.~\ref{fig:powerspectrum} show the poloidal and toroidal spectra of the ``turbulent" part of the magnetic field, $\mathbf{B}^T$, which is obtained by removing the contribution from the dynamo-like field:
\begin{equation}
\mathbf{B}^T \equiv \mathbf{B} - \mathbf{B}^D,
\label{eq:dyn}
\end{equation}
where $\mathbf{B}^D\equiv\langle \mathbf{B} \rangle_x$. While the turbulent and total spectra of the poloidal field are very similar (see panels $a$ and $c$, respectively), the turbulent spectra of the toroidal field (panel $d$) decrease substantially at  $0.2 \lesssim k\,2\pi v_{A,0}/\Omega_0\lesssim 1$ compared to the total toroidal spectra (panel $b$), maintaining its $\nu\approx 2$ behavior for $k\,2\pi v_{A,0}/\Omega_0 \gtrsim 1$. The $\nu\approx 5/3$ and 2 behaviors of the poloidal and toroidal components of the turbulent field is similar to the unstratified results from 3D MHD simulations of the MRI \citep[e.g.,][]{WalkerEtAl2016}, as well as the ones of 3D kinetic simulations \citep{KunzEtAl2016,BacchiniEtAl2022}.  
\begin{figure}
\centering
%\vspace{0.3cm}
 \hspace*{-0.cm}\includegraphics[width=1\linewidth]{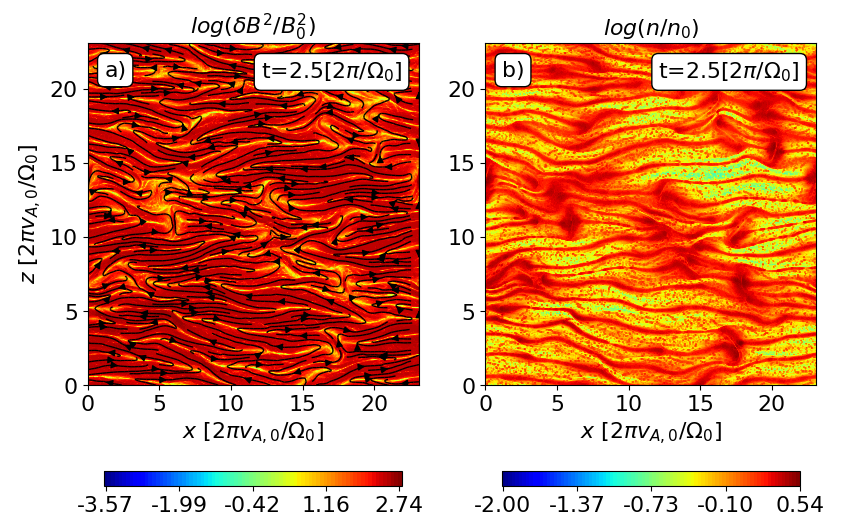}
% \vspace*{-1cm}
 \hspace*{-0.cm}\includegraphics[width=1\linewidth]{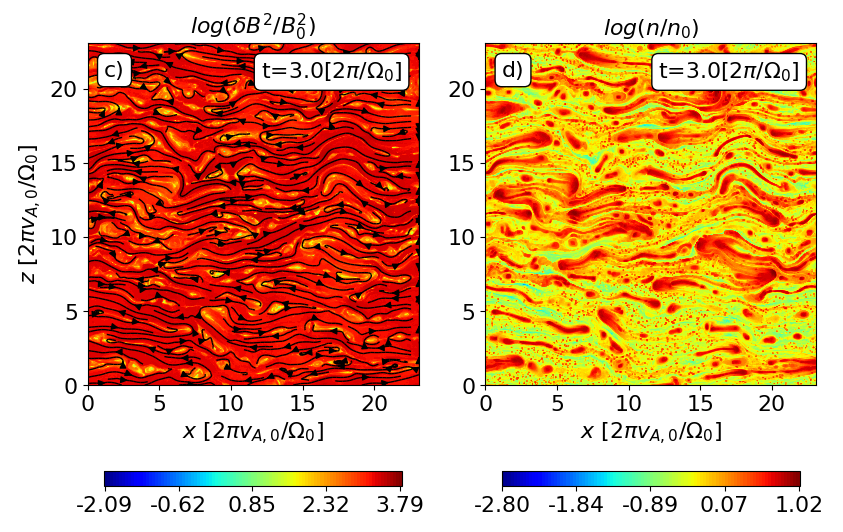}
% \hspace*{0.cm}\includegraphics[width=\linewidth]{fluc_fig64.png}
\caption{The squared magnetic fluctuations $\delta B^2$ (left) and the plasma density $n$ (right) for simulation UN2D-20 at $t=2.5$ and 3 [$2\pi/\Omega_0$]. The black arrows in the $\delta B^2$ panels show the total magnetic field projected on the $x-z$ plane.}
\label{fig:uns_snaps}
\end{figure}
\begin{figure}
    \centering
    \hspace*{-0.cm}\includegraphics[width=1\columnwidth]{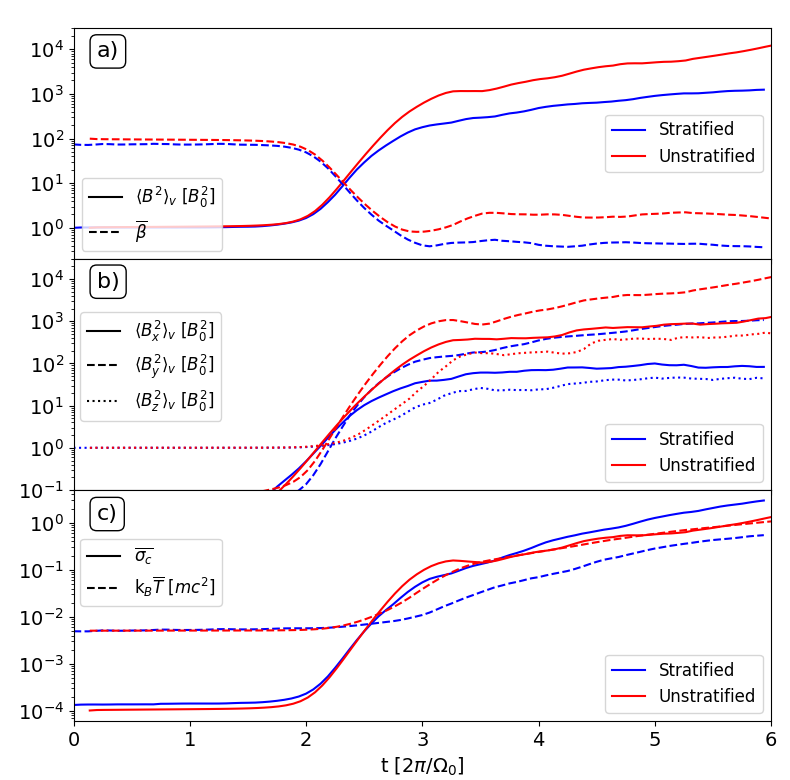}
    \caption{Plasma properties as a function of time $t$ for the disk region of the stratified run ST2D-20 (blue) and for the entire domain of the unstratified run UN2D-20 (red), respectively. Panel $a$ shows $\langle B^2 \rangle_v$ (solid) and $\overline{ \beta}$ (dashed). Panel $b$ shows the contributions to $\langle B^2 \rangle_v$ by the $x$ (solid), $y$ (dashed) and $z$ (dotted) components of the magnetic field. Panel $c$ shows $\overline{ \sigma}_c$ (solid) and $\overline{ T}$ (dashed).}
    \label{fig:volav_su}
\end{figure}

\subsection{Disk properties in 2D}
\label{sec:volume}

\noindent In this section we show the evolution of the average disk properties in our 2D runs, paying attention to the way these properties are affected by the presence of stratification and by the scale-separation ratio $\omega_{c,0}/\Omega_0$.\newline% Thus, we begin by providing a definition of the disk region, which is given by the condition 
%\noindent The occurrence of outflows and disk expansion observed in runs ST2D-20 and ST2D-3.5 has several effects on the propershties of the disk. In order to assess these effects, we compared run ST2D-20 with an unstratified simulation, run UN2D-20, where the initial conditions are the same as in the disk midplane of run ST2D-20. 
 %$|z| < H$, where $H(\overline{ T})$ $=(2k_B \overline{ T}/m)^{1/2}/\Omega_0$ is the scale height of a disk of temperature $\overline{ T}$ in hydrostatic equilibrium. Notice that the calculation of $\overline{ T}$ has to be done in the disk region, whose definition depends on $\overline{ T}$ itself through the inequality $|z| < H(\overline{ T})$, implying that $\overline{ T}$ is determined recursively.\newline

%\noindent Figure \ref{fig:uns_snaps} shows $\delta B^2$ and $n$ for run UN2D-20 at the same times used in Figure \ref{fig:strat_snaps} for run ST2D-20. At first glance, this figure suggests that the evolution of the MRI turbulence in run UN2D-20 is similar to the one in the disk of run ST2D-20, which we mark by the horizontal dotted lines in panels $b$, $d$ and $f$ of Fig. .... However, a more quantitative analysis using volume averaged quantities show that significant difference exist between the stratified and unstratified runs.
%%%
\noindent In order to assess the effect of stratification, we compare run ST2D-20 with the analogous unstratified run UN2D-20, with the same initial conditions as in the disk midplane of run ST2D-20. Figure \ref{fig:uns_snaps} shows $\delta B^2$ and $n$ for run UN2D-20 at the moment when nonlinear channel flows appear ($t=2.5$ $[2\pi/\Omega_0]$) and then when these channel flows have reconnected and broken into turbulence ($t=3$ $[2\pi/\Omega_0]$). At first glance, this figure suggests that the evolution of the MRI turbulence in run UN2D-20 is similar to the one in the disk region of run ST2D-20. However, the average plasma properties between these two runs differ substantially, as shown in Fig.~\ref{fig:volav_su}. Panel $a$ of Fig.~\ref{fig:volav_su} shows the evolution of $\langle B^2\rangle_v$ in the disk of run ST2D-20 (solid blue line) and in the whole volume of the analogous unstratified run UN2D-20 (solid red line). In both simulations there is an initial exponential growth regime that transitions to a much slower growth regime at $t \approx 3$ [$2\pi/\Omega_0$]. Also, the two runs show the lack of a complete magnetic field saturation.
%, consistent with the absence of reconnection of the toroidal component of the magnetic field shown by \cite{BacchiniEtAl2022}.
 However, at $t \gtrsim 3$ [$2\pi/\Omega_0$], at any given time the unstratified case reaches a $\langle B^2\rangle_v$ magnitude $\sim 5-10$ times larger than in the stratified case. This factor $\sim 5-10$ larger amplification applies similarly to the three components of the magnetic field, as can be seen from panel $b$ of Fig.~\ref{fig:volav_su}. \newline
 
\noindent Interestingly, the $B_y$ component in the unstratified case appears to be dominated by a large scale, dynamo-like component, similarly to what occurs in the stratified runs. This can be seen from panel $c$ of Figure \ref{fig:dynamo2d}, which shows $\langle B_y \rangle_x$ for run UN2D-20. We see that by $t=6$ $[2\pi/\Omega_0]$, $\langle B_y \rangle_x$ reaches an amplitude $\sim 100 B_0$, similar to the one of the total $B_y$ component, as shown by the dashed red line in panel $b$ of Fig.~\ref{fig:volav_su}. However, whereas the dynamo activity in the analogous stratified run ST2D-20 (shown in panel $a$ of Fig.~\ref{fig:dynamo2d}) produces a rather homogeneous $\langle B_y \rangle_x$ in the disk region ($|z| < H(\overline{ T})$, marked by the dashed black lines), the characteristic wavelength of the large scale $B_y$ field in the unstratified case is $\sim 4$ times smaller. We thus interprete the large scale $B_y$ field in the unstratified case as a growth in the wavelength of the MRI modes, being therefore of different nature compared to the larger scale $\langle B_y \rangle_x$ of the stratified runs.\newline
 
\noindent The time when the stratified simulation significantly slows down its growth ($t \approx 3$ [$2\pi/\Omega_0]$) coincides with the moment when stratification effects, such as outflows and disk expansion become important, as can be seen from Fig.~\ref{fig:strat_snaps}. Notice that this moment coincides with the time when the disk temperature starts increasing significantly, as we can see from the dashed-blue line in panel $c$ of Fig.~\ref{fig:volav_su}, showing a connection between energy dissipation and disk expansion and outflow generation.\newline
\begin{figure}
    \centering
\vspace{0.13cm}    \includegraphics[width=1.0\columnwidth]{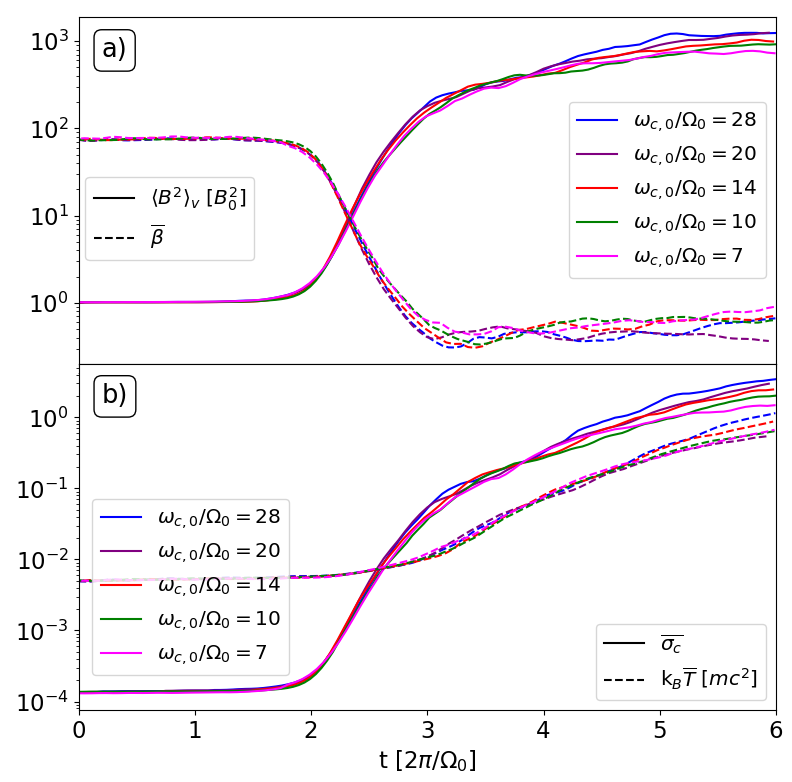}
    \caption{Plasma properties as a function of time $t$ for the disk region of 2D stratified runs using $\omega_{c,0}/\Omega_0=7$ (pink), 10 (green), 14 (red) and 20 (purple) and 28 (blue). Panel $a$ shows $\langle B^2 \rangle_v$ (solid) and $\overline{ \beta}$ (dashed). Panel $b$ shows $\overline{ \sigma_c}$ (solid) and $\overline{ T}$ (dashed).}
    \label{fig:Bamp_beta_mags}
\end{figure}
\noindent Despite the fact that the magnetic field is amplified less in the stratified case, the average cold sigma parameter $\overline{\sigma}_c$ ($\equiv \langle B^2\rangle_v/\langle 4\pi n m c^2\rangle_v$) is larger in the nonlinear regime of these runs. This is shown by the solid blue and solid red lines in panel $c$ of Fig.~\ref{fig:volav_su} for the stratified and unstratified cases, respectively. This can be explained by the decrease in the disk density $n$ due to its expansion in the stratified runs. Finally, in both cases the plasma beta $\overline{\beta}$ ($\equiv\langle 8\pi P \rangle_v/\langle B^2 \rangle_v$) reaches a nearly steady state regime for $t \gtrsim 3$ [$2\pi/\Omega_0]$, as shown by the dashed lines in panel $a$ of Fig.~\ref{fig:volav_su}. However, while $\overline{\beta}\sim 2$ in the unstratified case,  $\overline{\beta}\sim 0.4$ in the stratified case, which shows that stratification produces a disk that is magnetic-pressure supported, consistently with previous MHD stratified simulations \citep{BaiEtAl2013,SalvesenEtAl2016}.\newline

%\noindent Since the value of the scale-separation ratio, $\omega_{c,0}/\Omega_0$, in our simulations is much smaller than the one expected in realistic astrophysical settings, it is important to understand the effect of this parameter on the average properties of the disk. 
\noindent In order to explore the role of the scale-separation ratio $\omega_{c,0}/\Omega_0$ in our stratified runs, panels $a$ and $b$ of Figure \ref{fig:Bamp_beta_mags} show the quantities $\langle B^2\rangle_v$, $\overline{\beta}$, $\overline{\sigma}_c$, and $\overline{T}$ for stratified simulations with $\omega_{c,0}/\Omega_0=$ 7, 10, 14, 20 and 28. We see that increasing $\omega_{c,0}/\Omega_0$ produces a slight increase in $\langle B^2\rangle_v$ and $\overline{\sigma}_c$, not yet showing a clear convergence for the highest values of $\omega_{c,0}/\Omega_0$. (Note that the time origins of these simulations were slightly adjusted to align their exponential growth temporally, facilitating comparison). %However, when comparing different runs at the orbital time, they are assessed at slightly different evolutionary stages. To address this, Figure \ref{fig:Bamp_beta_Temp} presents $\langle B^2\rangle_v$, $\overline{\beta}$, and $\overline{\sigma}_c$ as a function of the average temperature in the disk, $\overline{T}$, for the same range of $\omega_{c,0}/\Omega_0$ values depicted in Figure \ref{fig:Bamp_beta_mags}. 
The evolutions of $\overline{T}$ and $\overline{\beta}$ exhibit some variations within a factor of $\sim 2$, but without showing a discernible dependence on $\omega_{c,0}/\Omega_0$.\newline

\begin{figure}
    \centering
    \hspace*{-0.cm}\includegraphics[width=1\columnwidth]{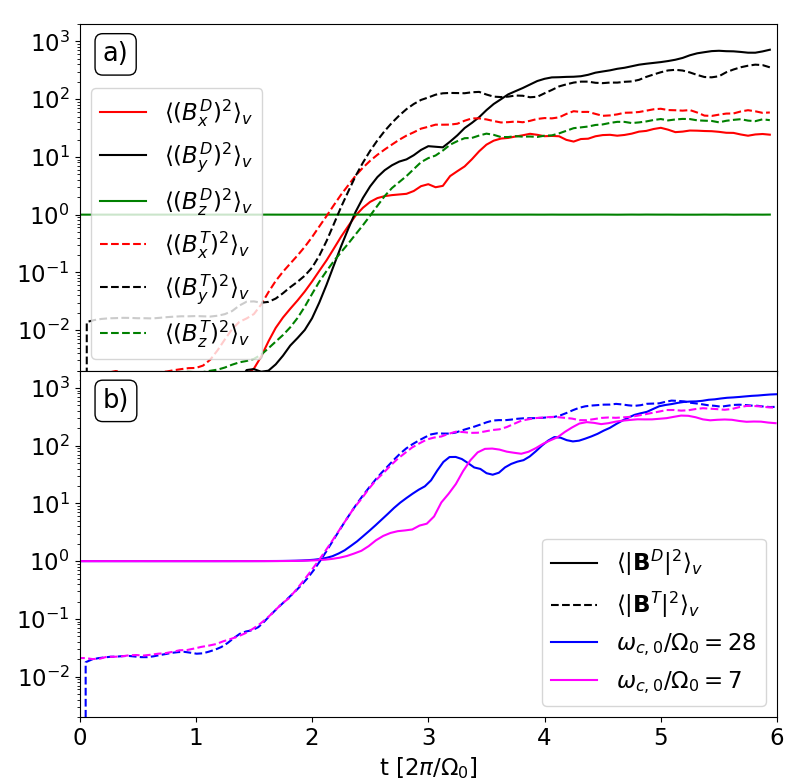}
    \caption{Panel $a$: the solid lines show $\langle(B_x^D)^2\rangle_v$ (red), $\langle(B_y^D)^2\rangle_v$ (black), and $\langle(B_z^D)^2\rangle_v$ (green), respectively, for run ST2D-20 ($\omega_{c,0}/\Omega_0=20$). The dashed lines show $\langle (B_x^T)^2 \rangle_v$ (red), $\langle (B_y^T)^2 \rangle_v$ (black) and $\langle (B_z^T)^2 \rangle_v$ (green) for the same run and in the same region. Panel $b$: the total energies in the dynamo-like field $\mathbf{B}^D$ (solid line) and in the turbulent field $\mathbf{B}^T$ (dashed line) for the runs ST2D-20 ($\omega_{c,0}/\Omega_0=28$; blue line) and ST2D-7 ($\omega_{c,0}/\Omega_0=7$; pink line).}
    \label{fig:dyn_comp}
\end{figure}

\noindent As discussed in \S \ref{sec:turbevol}, another important feature of the stratified 2D simulations is a dynamo-like action that produces a significant $B_y^D=\langle B_y \rangle_x$ field, as shown in panels $a$ and $b$ of Fig.~\ref{fig:dynamo2d}. The magnetic energy in the disk provided by the dynamo-like field $\mathbf{B}^D$ in run ST2D-20 is shown by the solid lines in the panel $a$ of Fig.~\ref{fig:dyn_comp}, where the red-solid, black-solid and green-solid lines show the contributions by the three components of $\mathbf{B}^D$: $\langle (B_x^D)^2\rangle_v$, $\langle (B_y^D)^2\rangle_v$ and $\langle (B_z^D)^2\rangle_v$, respectively. We see that the dynamo-like action within the disk is indeed dominated by the toroidal component of the magnetic field. Panel $a$ of Fig.~\ref{fig:dyn_comp} also shows in dashed lines the contribution to the magnetic energy provided by the three components of the turbulent magnetic field $\boldsymbol{B}^T$, which are averaged over the disk volume obtaining $\langle (B_x^T)^2 \rangle_v$, $\langle (B_y^T)^2 \rangle_v$ and $\langle (B_z^T)^2 \rangle_v$ (red-dashed, black-dashed and green-dashed lines, respectively). We see that the turbulent field is dominated by its toroidal component as well and contributes most of the magnetic energy in the disk from the triggering of the MRI turbulence at $t \approx 2$ [$2\pi/\Omega_0]$ until $t \approx 3.5$ [$2\pi/\Omega_0]$. After that, the toroidal component of the dynamo-like field $\langle\langle B_y \rangle_x^2\rangle_z$ becomes larger (by a factor of $\sim 2$) than the toroidal component of the turbulent field. \newline
\noindent Panel $b$ of Fig.~\ref{fig:dyn_comp} shows the total energies in the dynamo-like field $\mathbf{B}^D$ (blue-solid line) and in the turbulent field $\mathbf{B}^T$ (blue-dashed line) for run ST2D-20 ($\omega_{c,0}/\Omega_0=20$). We see that after $t \approx 4$ [$2\pi/\Omega_0]$ the energies in the dynamo and turbulent fields are comparable. Thus, in terms of the total magnetic energy, the dynamo-like and turbulent magnetic fields are roughly equally important after the initial period (of $\sim 1$ orbit after the triggering of the MRI) in which the turbulent magnetic energy dominates. This trend appears to not be significantly  affected by the scale-separation ratio $\omega_{c,0}/\Omega_0$. This is shown by the pink-solid and pink-dashed lines in panel $b$ of Fig.~\ref{fig:dyn_comp}, which show the contributions by, respectively, the turbulent and dynamo-like fields to the magnetic energy in the disk of run ST2D-7 ($\omega_{c,0}/\Omega_0=7$). We see that in this $\omega_{c,0}/\Omega_0=7$ run there is also an initial period of about $\sim 1$ orbit in which the turbulent field energy dominates, followed by a similar contribution to energy by the turbulent and dynamo-like fields.\newline

\noindent Thus, we have shown that disk stratification in 2D can change significantly the behavior of the MRI turbulence compared to the unstratified case. Besides producing significant outflows and a puffing up of the disk due to temperature increase, stratification makes the disk turbulence more magnetically dominated (smaller $\beta$) compared to what is shown in an analogous 2D unstratified simulation. Stratification also gives rise to a significant large scale dynamo-like activity, which contributes similarly to the magnetic energy in the nonlinear MRI stage as the turbulent field after $\sim 1$ orbit from the triggering of the MRI. 
We also found that increasing the scale-separation ratio produces slightly more magnetized disks, obtaining no complete convergence for the largest values of $\omega_{c,0}/\Omega_0$ used, consistent with the result obtained by \cite{BacchiniEtAl2022} who found that a scale separation ratio $\omega_{c,0}/\Omega_0\gtrsim 60$ is required for a complete convergence in the 2D case.\newline

\noindent In the next section we compare these 2D results with a 3D simulation showing that, although some differences appear, most of our 2D results are reasonably well reproduced in the 3D case.
\section{3D MRI turbulence}
\label{sec:3dmri}

\noindent In this section we present results from a 3D stratified simulation, run ST3D-3.5 ($\omega_{c,0}/\Omega_0=3.5$) and compare them with the analogous 2D stratified run ST2D-3.5.
\begin{figure}
    \centering
    \includegraphics[width=\columnwidth]{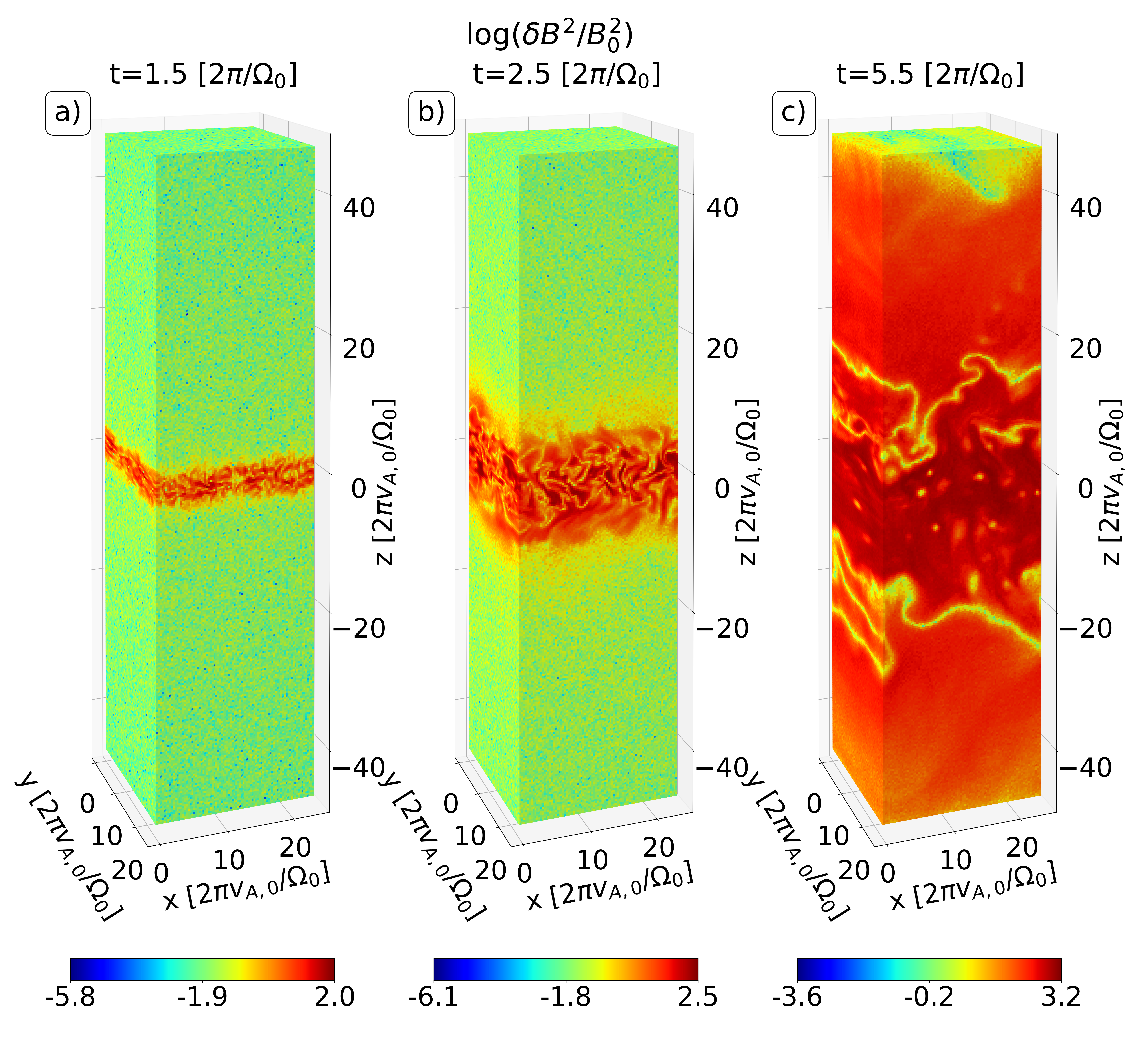}
    \caption{Panels $a$, $b$ and $c$ show $\delta B^2$ for simulation ST3D-3.5 at $t=1.5$, 2.5 and 4.5 [$2\pi/\Omega_0$], respectively.}
    \label{fig:3d1}
\end{figure}

\subsection{Turbulence properties in 2D vs 3D}
\label{sec:turbevol3d}
\begin{figure}
    \centering
    \includegraphics[width=\columnwidth]{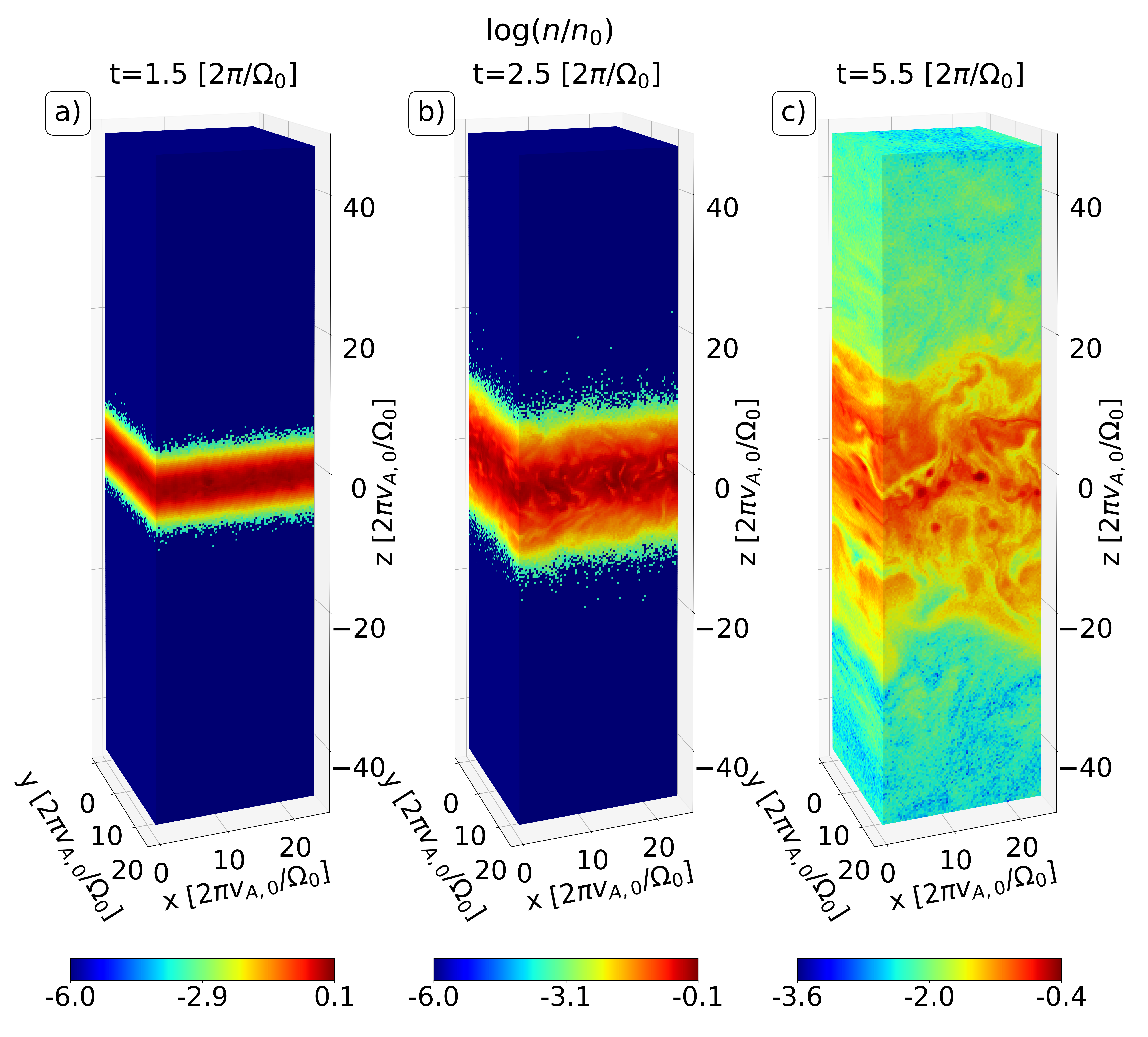}
    \caption{Panels $a$, $b$ and $c$ show $n$ for simulation ST3D-3.5 at $t=1.5$, 2.5 and 4.5 [$2\pi/\Omega_0$], respectively.}
    \label{fig:3d2}
\end{figure}

\noindent Figure \ref{fig:3d1} shows three snapshots of $\delta B^2$ for the stratified 3D run ST3D-3.5 at times $t=1.5$, 2.5 and 5.5 $[2\pi/\Omega_0]$. At the qualitative level, there are many similarities with the turbulence structure of the 2D runs. At $t=1.5$ $[2\pi/\Omega_0]$, nonlinear channel flows are present in $\delta B^2$, which look similar to the ones shown in panel $a$ of Fig.~\ref{fig:strat_snaps} for run ST2D-20. At $t=2.5$ $[2\pi/\Omega_0]$, the channel flows have already reconnected and broken into turbulence, with a significant increase in the disk thickness, similarly to what was shown for run ST2D-20 in panel $c$ of Fig.~\ref{fig:strat_snaps}. This trend continues at later times, as can be seen in panel $c$ of Fig.~\ref{fig:3d1}, which shows $\delta B^2$ at $t=5.5$ $[2\pi/\Omega_0]$. Figure \ref{fig:3d2} shows the same snapshots of Fig \ref{fig:3d1} but for the particle density $n$. At $t=1.5$ $[2\pi/\Omega_0]$ nonlinear channel flows are present in $n$, similarly to what is shown in panels $b$ of Fig.~\ref{fig:strat_snaps} for run ST2D-20. At $t=2.5$ and 4 $[2\pi/\Omega_0]$, a much more turbulent and progressively thicker disk is shown, as also shown for run ST2D-20 in panels $d$ and $f$ of Fig.~\ref{fig:strat_snaps}.\newline

\noindent Our 3D run also shows the action of a dynamo-like mechanism, as can be seen from panel $c$ of Fig.~\ref{fig:dynamo2d}, which shows $B_y$ averaged over the $x-y$ plane, $\langle B_y \rangle_{x-y}$, as a function of time $t$ and of the vertical coordinate $z$. We see that a net $\langle B_y \rangle_{x-y}$ field is formed, with an amplitude similar to the 2D cases shown in panels $a$ and $b$ of Fig.~\ref{fig:dynamo2d} (runs ST2D-20 and ST2D-3.5). However, while the dynamo-like field in 2D shows significant  time-variability and inhomogeneity along the $z$-coordinate, in 3D this field appears less variable and more homogeneous.\newline

% \noindent Figure \ref{fig:dynamo_decomp} shows the decomposition of the magnetic energy density from the mean magnetic field (solid), and the turbulent component (dotted) in the 3D run ST3D-3.5 (green) and the 2D run ST2D-3.5 (blue). In the 2D run the magnetic energy density is dominated by the fluctuating component, whereas in the 3D run the fluctuating component is $\sim 5$ times smaller than in 2D, being consistent with fact that 3D allow reconnection of the toroidal magnetic field component. However, in 3D the magnetic energy density at $t>3.5$ $[2\pi/\Omega_0]$ is dominated by the mean magnetic field.  \newline
%ccc% \noindent Further insights are gleaned from Figure \ref{fig:dynamo_decomp}, illustrating the magnetic energy density decomposition into its mean magnetic field component (solid line) and turbulent component (dotted line) for the 3D run ST3D-3.5 (green) and the 2D run ST2D-3.5 (blue). In the 2D run, the magnetic energy density is dominated by the fluctuating component, whereas in the 3D run, the fluctuating component is approximately five times smaller than its 2D counterpart. This finding aligns with the inherent reconnection capability of the toroidal magnetic field component facilitated by the three-dimensional nature of the simulation \citep{BacchiniEtAl2022}. However, in the 3D scenario, the magnetic energy density surpasses the fluctuating component and is dominated by the mean magnetic field for $t>3.5$ $[2\pi/\Omega_0]$.\newline

\begin{figure}
    \centering
    \hspace*{-0.cm} \includegraphics[width=1\columnwidth]{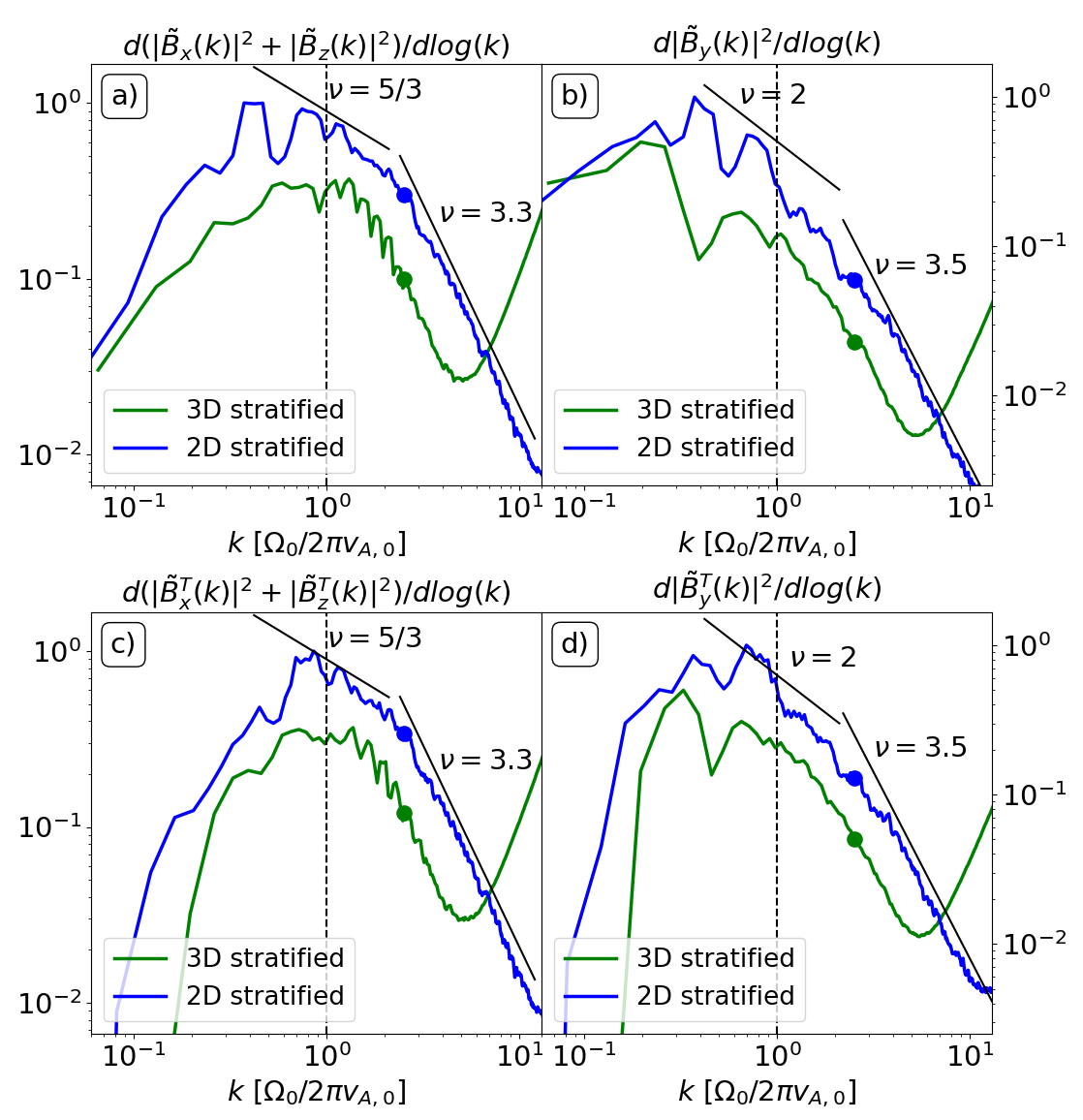}
    \caption{Panels $a$ and $b$ show the power spectra of the poloidal and toroidal components of the magnetic field, $d(|\tilde{B}_x(k)|^2+|\tilde{B}_z(k)|^2)/dln(k)$ and $d|\tilde{B}_y(k)|^2/dln(k)$, respectively, for the 2D and 3D stratified runs ST2D-3.5 and ST3D-3.5. Panels $c$ and $d$ show the same as in panels $a$ and $b$, respectively, but considering only the turbulent field $\mathbf{B}^T$.}
    \label{fig:powerspectrum2d3d}
\end{figure}

\noindent The behavior of the magnetic power spectrum seems to be quite similar in 2D and 3D. Panels $a$ and $b$ of Fig.~\ref{fig:powerspectrum2d3d} compare, respectively, the poloidal and toroidal magnetic spectra of the 2D and 3D runs ST3D-3.5 and ST2D-3.5 (blue and green lines, respectively). These runs share the same ratio $\omega_{c,0}/\Omega_0=3.5$, so that the effect of the scale-separation does not affect significantly the comparison. At sub-Larmor scales ($k\rho_l > 1$, where $k\rho_l = 1$ is marked by the colored dots on each line), we observe a magnetic spectrum with $\nu \approx 3.3$ (poloidal case) and $\nu \approx 3.5$ (toroidal case), for both types of runs. These $\nu \approx 3.3$ and 3.5 spectra are, however, steeper than the ones shown by the 2D runs with higher $\omega_{c,0}/\Omega_0$, showing that a minimum scale-separation ratio is necessary for correctly capturing the behavior of the sub-Larmor part of the spectra. Above Larmor scales ($k\rho_l < 1$), both runs show a poloidal and toroidal magnetic field spectra with $\nu$ close to $\nu\approx 5/3$ and $\nu\approx 2$, respectively. These $\nu\approx 5/3$ and  $\nu\approx 2$ behaviors are maintained when removing the dynamo-like field in both runs (which in 3D is defined as in Eq.~\ref{eq:dyn} but with $\mathbf{B}^D \equiv \langle \mathbf{B} \rangle_{x-y}$). This is shown in panels $c$ and $d$ of Fig.~\ref{fig:powerspectrum2d3d} where we show the poloidal and toroidal spectra of $\mathbf{B}^T$. The main effect of removing $\mathbf{B}^D$ is to substantially reduce the contribution of $k\,2\pi v_{A,0}/\Omega_0\lesssim 1$ to the toroidal part of the 2D and 3D spectra. In this way, the peak of the poloidal and toroidal spectra of $\mathbf{B}^T$ both in 2D and 3D approach $k\,2\pi v_{A,0}/\Omega_0\sim 1$ (although the toroidal part of the turbulent spectrum in 3D has its peak at wavelengths $\sim 3$ times larger than in 2D). This behavior of the $\mathbf{B}^T$ spectra in runs ST3D-3.5 and ST2D-3.5 above Larmor scales is similar to the ones shown in our stratified 2D runs with higher scale-separation ratio (Fig.~\ref{fig:powerspectrum}), as well as in previous unstratified MHD \citep{WalkerEtAl2016} and kinetic 3D simulations \citep{KunzEtAl2016,BacchiniEtAl2022}.

\subsection{Validation of assumptions}
\label{sec:valid}

\begin{figure}
    \centering
	    \includegraphics[width=1\columnwidth]{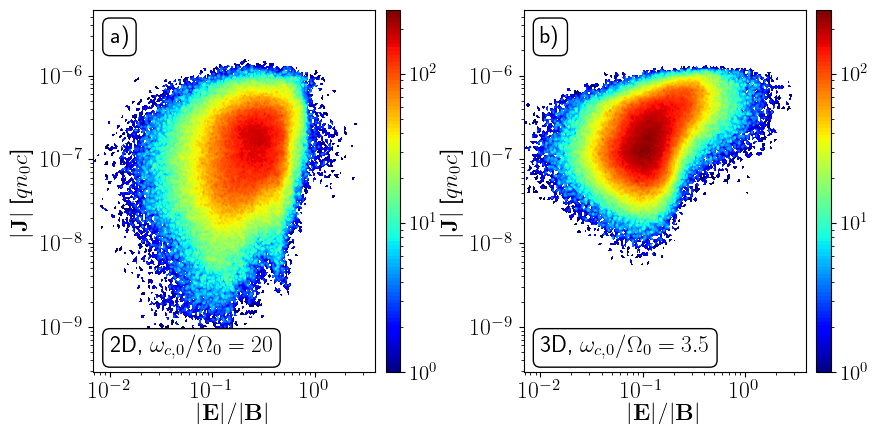}
    \caption{Panels a and b show the distribution of current density magnitude $|\mathbf{J}|$ and the $|\mathbf{E}|/|\mathbf{B}|$ ratio in the stratified run ST2D-20 and ST3D-3.5 at time $t=5$ [$2\pi/\Omega_0$], respectively.}
    \label{fig:current_EB}
\end{figure}

\noindent The fact that the peaks of the poloidal and toroidal spectra of $\mathbf{B}^T$ are close to $k\,2\pi v_{A,0}/\Omega_0\sim 1$ in 3D is important for the validation of the shearing coordinates approach used in this work. Indeed, since $\mathbf{B}^D$ only depends on $t$ and $z$, substracting this quantity from the total field does not change the power spectra of the magnetic field for $k_x$ and $k_y$ ($k_x = \hat{x}\cdot \mathbf{k}$ and $k_y = \hat{y}\cdot \mathbf{k}$, where $\mathbf{k}$ is the wave vector in Fourier space). Thus, the dominance of $k\,2\pi v_{A,0}/\Omega_0 \sim 1$ for poloidal and toroidal components of $\mathbf{B}^T$ implies that the dominant wavelength of the magnetic fluctuations along the $x$ and $y$ axis is given by $\sim 2\pi v_{A,0}/\Omega_0$. This is indeed one of the assumptions made in our implementation of the shearing coordinates approach (\S \ref{sec:shearingcoordinates}), which, combined with the condition $v_{A,0}/c \ll 1$, allowed us to drop the $y-$dependent terms in the field evolution equations \ref{eq:farad_comp} and \ref{eq:Amp_comp}, as well as to obtain the momentum and particle position evolution equations \ref{eq:p} and \ref{eq:dydt}. Notice that our shearing coordinates approach also assumes that the electric field in the MRI turbulence is either smaller or of the same order of the magnetic field ($|\mathbf{E}|/|\mathbf{B}|\lesssim 1$). To support this assumption, panels $a$ and $b$ of Fig.~\ref{fig:current_EB} shows as example the distribution of the electric current magnitude $|\mathbf{J}|$ and the $|\mathbf{E}|/|\mathbf{B}|$ ratio for runs ST2D-20 and ST3D-3.5 at time $t=5$ [$2\pi/\Omega_0$]. We see that in both cases the entire distribution satisfies $|\mathbf{E}|/|\mathbf{B}|\lesssim 3$, including the regions with the largest value of $|\mathbf{J}|$, which are expected to correspond to reconnecting current sheets. This shows that the assumptions made in \S \ref{sec:shearingcoordinates} to derive the plasma evolution equations in our shearing coordinates are fully supported by our obtained MRI turbulence behavior.

\subsection{Disk plasma properties in 2D vs 3D}
\label{sec:volume3d}
\begin{figure}
    \centering
    \hspace*{-0.cm} \includegraphics[width=1\columnwidth]{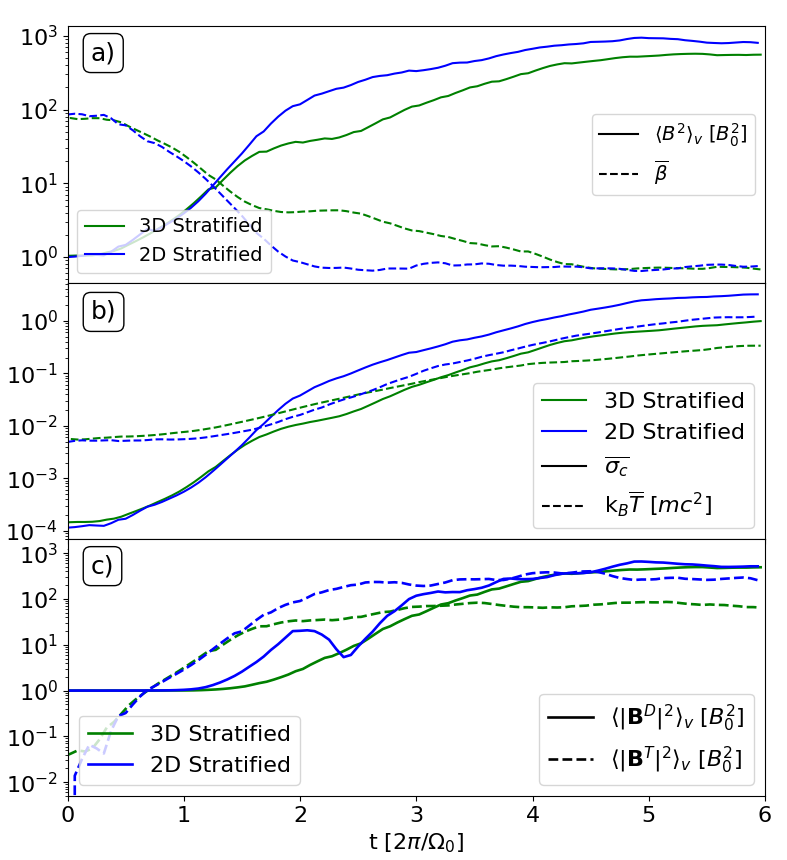}
    \caption{Average disk plasma properties as as a function of time $t$ for 2D (blue) and 3D (green) stratified runs ST2D-3.5 and ST3D-3.5, respectively. Panel $a$ shows $\langle B^2 \rangle_v$ (solid) and $\overline{ \beta}$ (dashed). Panel $b$ shows $\overline{\sigma}_c$ (solid) and $\overline{T}$ (dashed). Panel $c$ shows the magnetic energy densities in the turbulent field $|\mathbf{B}^T|^2$ (dashed) and in the dynamo-like field $|\mathbf{B}^D|^2$ (solid) in 2D and 3D.}
    \label{fig:saturation2d3d}
\end{figure}

\noindent In this section we compare the disk plasma properties evolution in 2D and 3D. Panel $a$ in Fig.~\ref{fig:saturation2d3d} shows the evolution of $\langle B^2 \rangle_v$ in the 2D and 3D runs ST2D-3.5 and ST3D-3.5. In both cases there is an initial exponential growth regime that evolves into a nonlinear regime with a much smaller growth rate at $t\approx 1.5$ $[2\pi/\Omega_0]$. Later, in the time interval $t\sim 1.5-3$ $[2\pi/\Omega_0]$, significant differences appear in the 2D and 3D cases, with the 3D run having a $\langle B^2\rangle_v$ amplitude $\sim 5$ times smaller. This significant difference in $\langle B^2\rangle_v$ produces a similar difference in $\overline{\sigma}_c$, as can be seen from the solid blue and green lines in panel $b$ of Fig.~\ref{fig:saturation2d3d}. This implies that in that time interval the disk expansion (and therefore its density) is about the same in the two simulations. This is consistent with the fact that their temperatures $\overline{ T}$ reach similar values, as shown by the dashed lines in panel $b$ of Fig.~\ref{fig:saturation2d3d}. Finally, consistently with the behaviors of $\overline{T }$ and $\overline{\sigma}_c$, $\overline{\beta}$ is $\sim 5$ times larger in the 3D case during the time period $t\sim 1.5-3.5$ $[2\pi/\Omega_0]$. Later, when $t\gtrsim 3$ $[2\pi/\Omega_0]$ there is a transition towards a state in which the amplitudes of $\langle B^2\rangle_v$ in 2D and 3D tend to give more similar values, which also tends to produce similar values of $\overline{\beta}$. Indeed, for $t\gtrsim 4$ $[2\pi/\Omega_0]$, $\overline{\beta} \approx 0.5$ in both runs while $\langle B^2 \rangle_v$ is only a factor of $\sim 2$ larger in the 2D case.\newline

\noindent The smaller magnetic amplification shown by the 3D run in the time interval $t\sim 1.5-4$ $[2\pi/\Omega_0]$ is consistent with recent unstratified PIC simulations of the MRI that show that using 3D runs is important to allow efficient reconnection of the toroidal magnetic field component \citep{BacchiniEtAl2022}. By the end of the simulations, however, the 2D and 3D magnetic energies only differ by a factor of $\sim 2$. This can be explained by the growing importance of the dynamo-like field in the stratified 2D and 3D runs, which evolves very similarly in these two types of runs. The progressively growing importance of the dynamo-like field in 2D can be seen from panel $c$ of Fig.~\ref{fig:saturation2d3d}, which shows that in run ST2D-3.5, $|\mathbf{B}^D|^2$ (solid-blue line) starts smaller than the turbulent part of the magnetic energy density $|\mathbf{B}^T|^2$ (dotted-blue line) for $t\lesssim 4$ $[2\pi/\Omega_0]$, but afterwards it becomes comparable to $|\mathbf{B}^T|^2$. This is indeed consistent with what was shown for 2D runs with larger scale-separation ratios in Fig.~\ref{fig:dyn_comp}. In the 3D run ST3D-3.5 this increase in the dynamo-like field importance is even more significant, since $|\mathbf{B}^D|^2$ (solid-green line) becomes $\sim 5$ times larger than $|\mathbf{B}^T|^2$ (dotted-green line) at $t\gtrsim 4$ $[2\pi/\Omega_0]$, given that 3D runs dissipate $|\mathbf{B}^T|^2$ more efficiently via reconnection. Since $|\mathbf{B}^D|^2$ has essentially the same values in 2D and 3D, it is thus expected that, by the end of the simulations, $\langle B^2 \rangle_v$ only differs by a factor of $\sim 2$ between the 2D and 3D cases.\newline
\noindent In \S \ref{sec:effectivevisc2} we show that the similitude between 2D and 3D by the end of the runs is also reproduced when analyzing the MRI-driven effective viscosity. 

\section{Effective viscosity}
\label{sec:effectivevisc}
In this section we analyze the effective disk viscosity caused by the MRI turbulence. This viscosity is quantified making used of the $\alpha$ parameter \citep{ShakuraEtAl1973}, defined as the $xy$ component of the plasma stress tensor $T_{xy}$, normalized by the plasma pressure, $\alpha = T_{xy}/P$. This stress tensor component $T_{xy}$ has three contributions: the Maxwell stress $M_{xy}=-B_xB_y/4\pi$, the Reynolds stress $R_{xy} = mnV_xV_y$, where $\boldsymbol{V}=(V_x,V_y,V_z)$ is the fluid velocity, and the anisotropic stress $A_{xy} = - (P_{\perp}-P_{\parallel})B_xB_y/B^2$, where $P_{\perp}$ and $P_{\parallel}$ are the plasma pressures perpendicular and parallel to the local magnetic field. Notice that, even though in the calculation of $R_{xy}$ we assume non-relativistic fluid velocities, in our simulations individual particles can still acquire relativistic velocities. Thus $\boldsymbol{V}$ is calculated as $\boldsymbol{V} = \langle \boldsymbol{p} \rangle_{p}/m\langle \gamma \rangle_{p}$, where $\boldsymbol{p}$ and $\gamma$ are the momenta and Lorentz factors of the particles in a given fluid element and $\langle \, \rangle_{p}$ denotes an average over those particles. In this way we ensure that the fluid velocity $\boldsymbol{V}$ corresponds to the velocity of the reference frame where the average particles momentum within a fluid element vanishes.

\subsection{Effect of stratification and $\omega_{c,0}/\Omega_0$ on viscosity}
\label{sec:effectivevisc1}
\begin{figure}
\centering
\includegraphics[width=1\columnwidth]{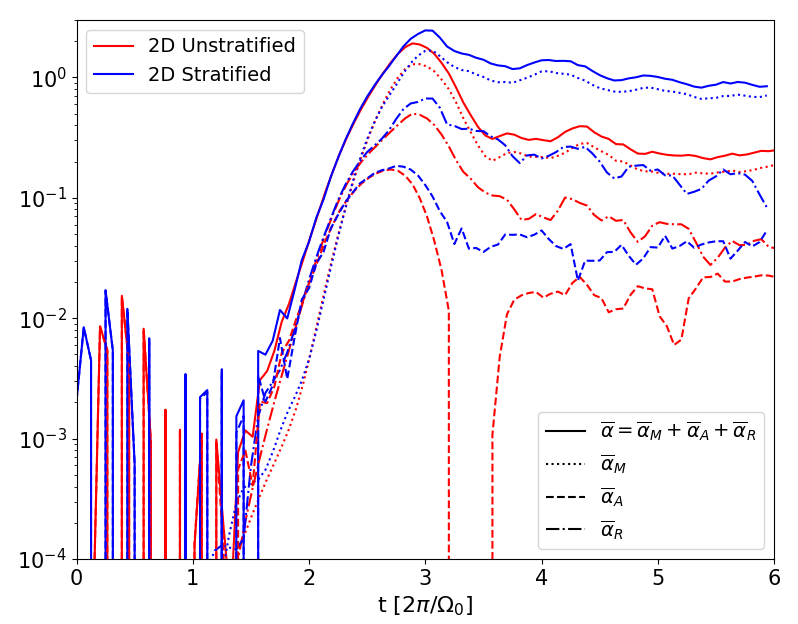}
\caption{The solid lines show the evolution of $\overline{\alpha}$ in the stratified run ST2D-20 (blue) and the unstratified run UN2D-20 (red). The contributions from the Maxwell, Reynolds, and anisotropic stresses ($\overline{\alpha}_M$, $\overline{\alpha}_R$ and $\overline{\alpha}_A$, respectively) are also shown by the dotted, dash-dotted and dashed lines, respectively.}
    \label{fig:alpha_comp}
\end{figure}

\noindent Fig.~\ref{fig:alpha_comp} shows in solid blue line the time evolution of the average parameter $\overline{\alpha}$ ($\equiv \langle T_{xy}\rangle_v/\langle P\rangle_v$) for run ST2D-20, along with the contributions from the Maxwell, Reynolds and anisotropic stresses: $\overline{\alpha}_M$ ($\equiv \langle M_{xy}\rangle_v/\langle P\rangle_v$; dotted blue line), $\overline{\alpha}_R$ ($\equiv \langle R_{xy}\rangle_v/\langle P\rangle_v$; dot-dashed blue line) and $\overline{\alpha}_A$ ($\equiv \langle A_{xy}\rangle_v/\langle P\rangle_v$; dashed line), respectively. We see that $\overline{\alpha}$ reaches a saturated value of $\overline{\alpha} \sim 1$, which is dominated by the Maxwell stress, with the contributions to $\overline{\alpha}$ following the ordering $\overline{\alpha}_M > \overline{\alpha}_R > \overline{\alpha}_A$. The fact that $\overline{\alpha}_M\sim 1$ is consistent with the dominance of magnetic pressure compared to particle pressure ($\overline{\beta}\lesssim 1$) seen in panel $a$ of Fig.~\ref{fig:volav_su} for the same run ST2D-20. We compare these results with the ones of the analogous unstratified run UN2D-20, where we find a similar ordering of the contributions to $\overline{\alpha}$, $\overline{\alpha}_M > \overline{\alpha}_R > \overline{\alpha}_A$, but with a $\sim 4$ times smaller $\overline{\alpha}_M$. This difference is consistent with the $\sim 4$ times larger $\overline{\beta}$ obtained in the unstratified run, implying a significant effect of stratification on the disk viscosity in collisionless studies of the MRI. Notably, the importance of $\overline{\alpha}_R$ in our results seems to contradicts previous unstratified kinetic studies \citep{KunzEtAl2016,BacchiniEtAl2022}, but are in line with the findings from MHD simulations \citep{BaiEtAl2013,SalvesenEtAl2016}. \newline
\begin{figure}
    \centering
    \includegraphics[width=1.0\columnwidth]{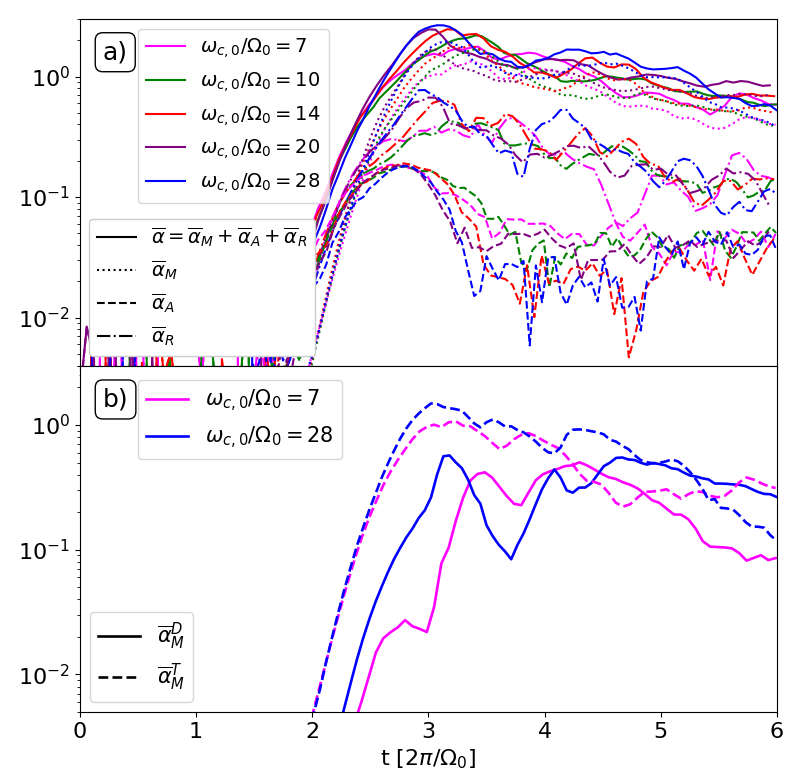}
    \caption{Panel $a$ shows in solid lines the evolution of $\overline{\alpha}$ in 2D runs with $\omega_{c,0}/\Omega_0=7$ (ST2D-7; pink), $\omega_{c,0}/\Omega_0=10$ (ST2D-10; green), 14 (ST2D-14; red), 20 (ST2D-20; purple) and 28 (ST2D-28; blue). The corresponding contributions from the Maxwell, Reynolds, and anisotropic stresses ($\overline{\alpha}_M$, $\overline{\alpha}_R$ and $\overline{\alpha}_A$, respectively) are also shown by the dotted, dash-dotted and dashed lines, respectively. Panel $b$ shows in blue lines the contribution to $\overline{\alpha}_M$ by the turbulent field $\mathbf{B}^T$ (dashed lines) and dynamo-like fields $\mathbf{B}^D$ (solid lines) in run ST2D-28 run ($\omega_{c,0}/\Omega_0=28$), which we name $\overline{\alpha}_M^T$ and $\overline{\alpha}_M^D$, respectively. $\overline{\alpha}_M^T$ and $\overline{\alpha}_M^D$ for run ST2D-7 ($\omega_{c,0}/\Omega_0=7$) are shown by dashed-pink and solid-pink lines, respectively.}
    \label{fig:alpha_sepa}
\end{figure}
\begin{figure}
    \centering
    \vspace{-0.05 cm}
    \includegraphics[width=.995\columnwidth]{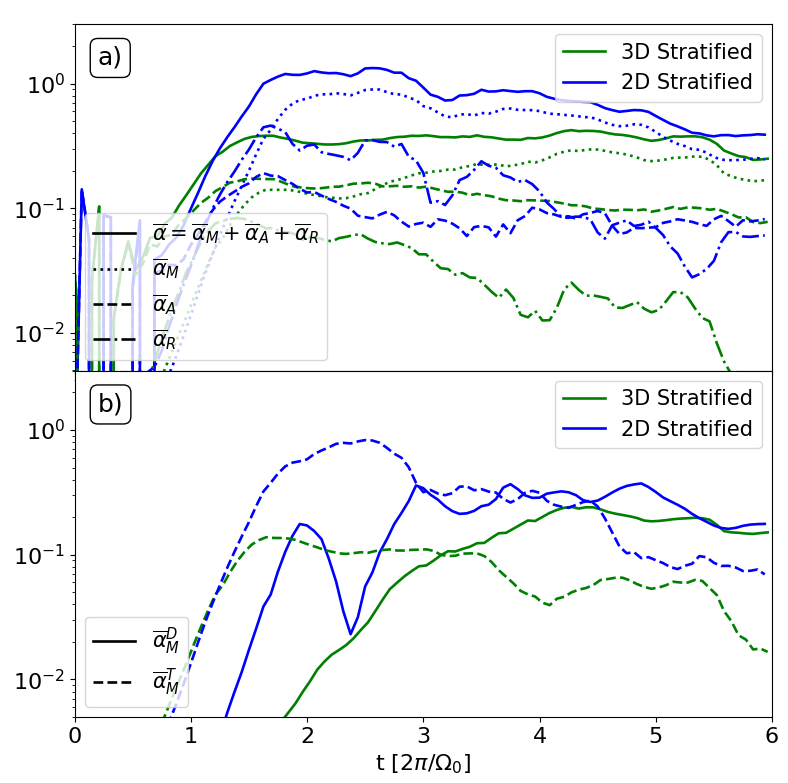}
    \caption{Panel $a$ shows in solid lines the evolution of $\overline{\alpha}$ in the 3D run ST3D-3.5 (green) and the 2D run ST2D-3.5 (blue). Their contributions from the Maxwell, Reynolds, and anisotropic stresses ($\overline{\alpha}_M$, $\overline{\alpha}_R$ and $\overline{\alpha}_A$, respectively) are also shown by the dotted, dash-dotted and dashed lines, respectively. Panel $b$ shows the contributions of the dynamo-like magnetic field (solid) and the turbulent magnetic field (dashed) to the Maxwell stress, $\overline{\alpha}_M^D$ and $\overline{\alpha}_M^T$, in the 3D run ST3D-3.5 (green) and the 2D run ST2D-3.5 (blue).}
    \label{fig:alphas2d3d}
\end{figure}

\noindent We also measured the effect of $\omega_{c,0}/\Omega_0$ on the behavior of $\overline{\alpha}_M$, $\overline{\alpha}_R$, $\overline{\alpha}_A$ and the total $\overline{\alpha}$, which is done in panel $a$ of Fig.~\ref{fig:alpha_sepa}. We see that, although $\overline{\alpha}$ fluctuates by factors of order unity, there is no discernible dependence of this quantity on $\omega_c/\Omega_0$, implying that the scale separation used in our 2D runs appears to be large enough to accurately capture the behavior of the MRI-driven viscosity. The blue lines in panel $b$ of Fig.~\ref{fig:alpha_sepa} also compare the contribution to $\overline{\alpha}_M$ by the turbulent field $\mathbf{B}^T$ (dashed lines) and dynamo-like fields $\mathbf{B}^D$ (solid lines) in the ST2D-28 run ($\omega_{c,0}/\Omega_0=28$), which we name $\overline{\alpha}_M^T$ and $\overline{\alpha}_M^D$, respectively. We see that $\overline{\alpha}_M^T$ dominates until $t\sim 4$ $[2\pi/\Omega_0]$. After that moment $\overline{\alpha}_M^T$ and $\overline{\alpha}_M^D$ are comparable, with fluctuating differences of a factor $\sim 2-3$). A similar behavior is obtained for $\overline{\alpha}_M^T$ and $\overline{\alpha}_M^D$ in run ST2D-7 ($\omega_{c,0}/\Omega_0=7$), which are shown by dashed-pink and solid-pink lines, respectively. This implies that no clear effect of the scale-separation ratio of $\overline{\alpha}_M^T$ and $\overline{\alpha}_M^D$ is observed in our simulations. The fact that these quantities become comparable after $t\sim 4$ $[2\pi/\Omega_0]$ is in line with the behaviors of $|\mathbf{B}^T|^2$ and $|\mathbf{B}^D|^2$ for runs ST2D-28 and ST2D-7, which also become comparable in the same time period, as shown in panel $b$ of Fig.~\ref{fig:dyn_comp}.

\subsection{Viscosity in 2D vs 3D}
\label{sec:effectivevisc2}
\noindent Panel $a$ of Fig.~\ref{fig:alphas2d3d} compares the effective viscosities of the 3D and 2D runs ST3D-3.5 and ST2D-3.5, respectively. We see that for $t\gtrsim 1.5$ $[2\pi/\Omega_0]$, the viscosity of the 3D run has a nearly steady value of $\overline{\alpha} \approx 0.5$. For $t\sim 1.5-3.5$ $[2\pi/\Omega_0]$, the 3D $\overline{\alpha}$ is $\sim 3-4$ times smaller than in the 2D case, while for $t\gtrsim 3.5$ $[2\pi/\Omega_0]$ the $\overline{\alpha}$ of the 2D and 3D runs become more similar, differing by a maximum factor of $\sim 2$. The time dependence of the difference between the 2D and 3D values of $\overline{\alpha}$ is consistent with the fact that, initially, the 3D $\overline{ \beta}$ is $\sim 3-5$ times larger than in 2D, with a subsequent period at $t\gtrsim 3.5$ $[2\pi/\Omega_0]$ in which both $\overline{\alpha}$'s acquire essentially the same value, as shown by the dashed blue (2D) and green (3D) lines in panel $a$ of Fig.~\ref{fig:saturation2d3d}.\newline
%\begin{figure}
%\centering
%\includegraphics[width=\columnwidth]{Maxwell_decom.png}
%\caption{The solid lines show the evolution of $\overline{\alpha}_M$ in the 3D run ST3D-3.5 (green) and the 2D run ST2D-3.5 (blue). Their contributions from the mean and fluctuating magnetic field ($\overline{\alpha_M^{mean}}$ and $\overline{\alpha_M^{fluc}}$) are also shown by the dotted and dashed lines.}
%\label{fig:Maxwell_decomp}
%\end{figure}

\noindent %Furthermore, the 3D runs reveal a lower intensity of turbulent magnetic field compared to the 2D run, along with a higher mean magnetic field, as depicted in Figure \ref{fig:dynamo_decomp}. 
These results reinforce the idea that, when the dynamo-like field becomes either dominant (3D) or comparable to the turbulent field (2D) at $t\gtrsim 3.5$ $[2\pi/\Omega_0]$, the 2D and 3D runs produce fairly similar results, which include the value of the  (Maxwell stress-dominated) disk viscosity. When that happens, $\overline{\alpha}_M$ itself is significantly affected by the dynamo-like field. This can be seen from panel $b$ of Figure \ref{fig:alphas2d3d}, which shows the contributions of the dynamo-like magnetic field (solid) and the turbulent magnetic field (dashed) to the Maxwell stress, $\overline{\alpha}_M^D$ and $\overline{\alpha}_M^T$, in the 3D run ST3D-3.5 (green) and the 2D run ST2D-3.5 (blue). At $t>3.5$ $[2\pi/\Omega_0]$, the 3D run exhibits a greater contribution to the Maxwell stress attributed to the dynamo-like field (by a factor $\sim 5$). This dominant contribution to the viscosity by the large scale dynamo-like field is in qualitative agreement with the 3D MHD simulations of \cite{BaiEtAl2013} in the case of $\beta_0=100$. Conversely, in the 2D run ST2D-3.5 the dynamo-like contribution to the viscosity becomes comparable to the one of the turbulent field after $t \sim 3.5$ $[2\pi/\Omega_0]$, with some dominance of the former after $t \sim 4.5$ $[2\pi/\Omega_0]$ by a factor of $\sim 2-3$. This is in line with results shown for the 2D runs ST2D-28 and ST2D-7 ($\omega_{c,0}/\Omega_0=28$ and 7, respectively), for which $\overline{\alpha}_M^D$ and $\overline{\alpha}_M^T$ were comparable after $t \sim 4$ $[2\pi/\Omega_0]$, with no discernible dependence on $\omega_{c,0}/\Omega_0$.\newline

\noindent Both our 2D and 3D runs give rise to an anisotropic stress that is subdominant compared to the Maxwell stress, although the former is larger than the Reynolds stress in the 3D run, which is the contrary to what occurs in 2D, suggesting that 3D effects would tend to suppress the fluid velocities that give rise to the Reynolds stress. 
%\begin{figure*}
%    \centering
%     \includegraphics[width=2\columnwidth]{fig20.png}
%     \vspace*{-2.7cm}\caption{Panel $a$ shows the magnitude of the plasma current density, $|\boldsymbol{J}|$, for run ST2D-20 in its nonlinear stage ($t \approx 5$ $[2\pi/\Omega_0]$). Panel $b$ shows the heating rate per unit volume of the plasma at the same moment.}
%    \label{fig:heating}
%\end{figure*}

\subsection{Pressure anisotropy behavior}
\label{sec:effectivevisc3}
\begin{figure}
\centering
\vspace{0cm}
\includegraphics[width=1\columnwidth]{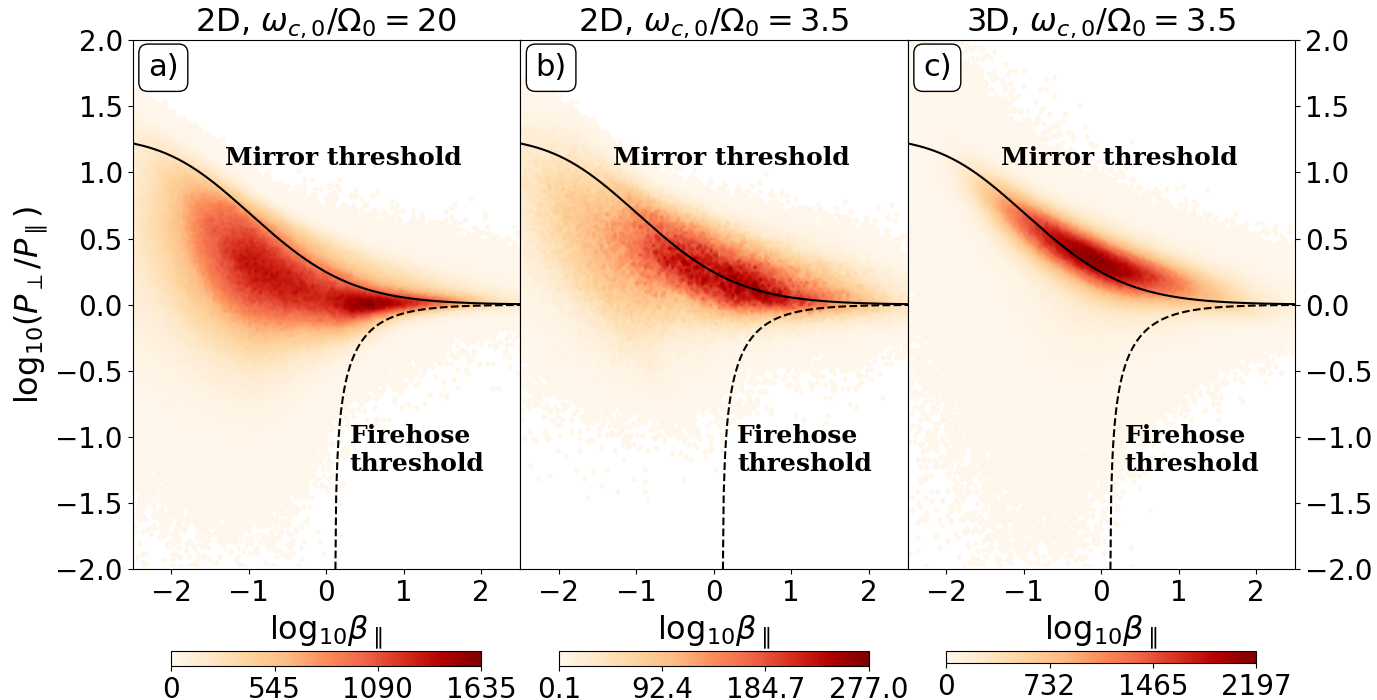}
\caption{Panels $a$, $b$ and $c$ show the distributions of plasma anisotropy $P_{\perp}/P_{\parallel}$ and $\beta_{\parallel}$ in the disk of the 2D run ST2D-20 during $t = 3.5-4.5$ $[2\pi/\Omega_0]$, the 2D run ST2D-3.5 during $t = 2.5-3.5$ $[2\pi/\Omega_0]$ and the 3D run ST3D-3.5 during $t = 2-3$ $[2\pi/\Omega_0]$, respectively. The three cases are compared with a threshold for the growth of unstable mirror modes (black line) and firehose modes (dashed line) obtained from linear Vlasov theory (Hellinger et al 2006).}
    \label{fig:aniso}
\end{figure}

\noindent The very small contribution of $\overline{\alpha}_A$ to the total effective viscosity in our 2D and 3D stratified runs seems to contradict previous kinetic simulation studies that suggest that the anisotropic stress can be as important as Maxwell stress in collisionless disks \citep[e.g.,][]{KunzEtAl2016}. This discrepancy, however, appears to be mainly due to the small $\beta$ regime reached in the nonlinear state of our simulations. To demonstrate this point, panel $a$ of Fig.~\ref{fig:aniso} shows the distribution of plasma anisotropy $P_{\perp}/P_{\parallel}$ and $\beta_{\parallel}$ in the disk of run ST2D-20 during a time interval $t = 3.5-4.5$ $[2\pi/\Omega_0]$, and compares it with a threshold for the growth of unstable mirror modes (black line) obtained from linear Vlasov theory (Hellinger et al 2006):
\begin{equation}
\frac{P_{\perp}}{P_{\parallel}} = 1 + \frac{0.77}{(\beta_{\parallel}-0.016)^{0.76}}.
\label{eq:threshold}
\end{equation}
We see that in most cases $P_{\perp}/P_{\parallel}$ tends to be larger than unity and limited by the mirror threshold. As an estimate of the upper limit for the expected importance of $\alpha_A$, one can compute the ratio $\langle\alpha_A/\alpha_M\rangle_v$ assuming that the pressure anisotropy of the plasma is given by Eq.~\ref{eq:threshold}. In that case we would have
\begin{equation}
\Big\langle\frac{\alpha_A}{\alpha_M}\Big\rangle_v =\Big\langle \frac{(P_{\perp} - P_{\parallel})}{P_{\parallel}}\frac{\beta_{\parallel}}{2}\Big\rangle_v \lesssim 0.4\langle\beta_{\parallel}^{0.24}\rangle_v\sim 0.4\overline{\beta}_{\parallel}^{0.24},  
\label{eq:ratioalphas}
\end{equation} 
where we have applied Eq.~\ref{eq:threshold} in the limit $\beta_{\parallel} \gg 0.016$ ($\overline{\beta}_{\parallel} \approx 0.4$ for $t \gtrsim 3$ [$2\pi/\Omega_0$], as can be seen from the $\overline{\beta}_{\parallel}$ evolution for run ST2D-20 shown in Fig.~\ref{fig:volav_su}). Thus, using $\overline{\beta}_{\parallel} \approx 0.4$, we obtain $\langle\alpha_A/\alpha_M\rangle_v \lesssim 0.3$. This upper limit is consistent with the fact that $\overline{\alpha}_A$ is much smaller (by a factor of $\sim 10$) than $\overline{\alpha}_M$ in run ST2D-20, as shown, respectively, by the dashed blue and dotted blue lines in Fig.~\ref{fig:alpha_comp}. Notice that in a hypothetical case in which $\overline{\beta}_{\parallel} \sim 100$ \citep[e.g., as in][]{KunzEtAl2016}, Eq.~\ref{eq:ratioalphas} would predict comparable contributions from the anisotropic and Maxwell stress with $\overline{\alpha}_A \sim \overline{\alpha}_M$. \newline

\noindent Panel $b$ of Fig.~\ref{fig:aniso} shows the same as panel $a$ but for the 2D run ST2D-3.5. We see that $P_{\perp}/P_{\parallel}$ is somewhat larger in the case of run ST2D-3.5 for a given $\beta_{\parallel}$. The larger value of $P_{\perp}/P_{\parallel}$ is consistent with the smaller scale-separation ratio, as shown by previous PIC simulation studies of the mirror instability driven by a growing background magnetic field \cite[see, e.g.,][]{LeyEtAl2023}. However, the distribution of $P_{\perp}/P_{\parallel}$ and $\beta_{\parallel}$ in the disk of run ST2D-3.5 still follows reasonably well the threshold for the growth of mirror modes presented in Eq.~\ref{eq:threshold}, consistently with the essentially absent effect of scale-separation on the dominance of $\overline{\alpha}_M$ in our runs. Panel $c$ of Fig.~\ref{fig:aniso} shows the behavior for $P_{\perp}/P_{\parallel}$ and $\beta_{\parallel}$ in the 3D run ST3D-3.5. We see that the pressure anisotropy behaves similarly in the runs ST2D-3.5 and ST3D-3.5, in agreement with the small contribution of $\overline{\alpha}_A$ to the effective viscosity in the 3D case.\newline

\noindent In summary, our 2D and 3D runs give a (Maxwell stress dominated) $\overline{\alpha}$ with values between  $\sim 0.5$ (3D) and $\sim 1$ (2D), with a progressively similar behavior of the 2D and 3D runs as the dynamo-like field becomes dominant ($t \gtrsim 3.5$ $[2\pi/\Omega_0]$). In this dynamo-dominated regime, $\overline{\alpha}$ is expected to be mainly produced by the dynamo-like field. Interestingly, this viscosity behavior is very similar to the one obtained from 3D MHD simulations of stratified disk with net vertical field and initial $\beta = 100$ \citep{SalvesenEtAl2016}. 

\section{Particle acceleration}
\label{sec:heataccel}

\noindent Our stratified MRI simulations show significant particle acceleration. In this section, we show that the acceleration efficiency grows as the disk temperature and the scale-separation ratio $\omega_{c,0}/\Omega_0$ increase. Well developed nonthermal tails are observed mainly in our 2D runs, due to their relatively large scale-separation ratio. 

\subsection{Spectrum evolution in 2D}
\label{sec:accel}
\begin{figure}
    \centering
    \includegraphics[width=.95\columnwidth]{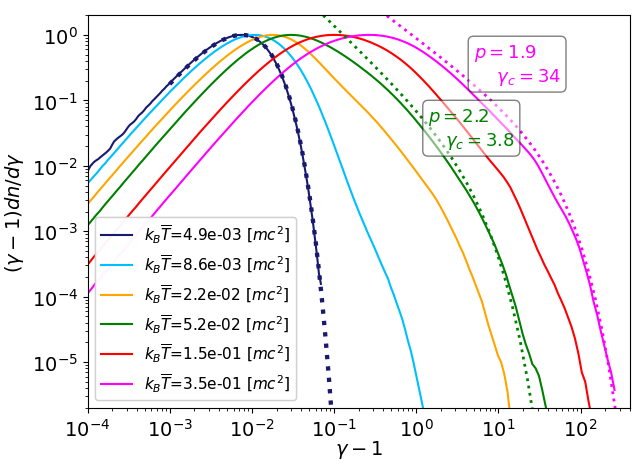}
    \caption{The evolution of the particle energy distribution for our fiducial run ST2D-20. Different colors represent different disk temperatures, and the dotted lines correspond to fits to the nonthermal tails using Eq.~\ref{eq:powlaw}.}
    \label{fig:spect_evol}
\end{figure}
%\noindent Panel $a$ of Fig \ref{fig:heating} shows the magnitude of the plasma current density, $|\boldsymbol{J}|$, for run ST2D-20 in its nonlinear stage ($t \approx 5$ $[2\pi/\Omega_0]$), and panel $b$ shows the distribution of simulation points in terms of their $|\boldsymbol{J}|$ and heating rate per particle at the same moment and in the disk region (limited by the horizontal dashed lines).\footnote{The heating rate is obtained by calculating the work done by all the forces that act on the particles, which are described by Eq.~\ref{eq:p}.} We see that the regions where $|\boldsymbol{J}|$ is the largest correspond to the locations where the energy gain per particle is the largest. This suggests that reconnecting current sheets contribute significantly to the acceleration of the most energetic particles in our simulations. \newline
\noindent The evolution of the particle spectrum, $dn/d\gamma$, calculated in the disk of run ST2D-20 is shown in Fig.~\ref{fig:spect_evol}, where $\gamma$ is the particle Lorentz factor. The spectra are shown for different values of the disk temperature $\overline{ T}$, instead of at different times. (This allows us to compare spectra from different simulations, removing the fact that different runs may take different times to trigger the MRI and/or to heat the plasma). As the plasma temperature increases, their spectra develop a nonthermal tail that can be approximately described as a power-law with an exponential cut-off,
\begin{equation}
\frac{dn}{d\gamma} \propto (\gamma-1)^{-p}e^{-\gamma/\gamma_{\textrm{c}}},
\label{eq:powlaw}
\end{equation}
where $p$ and $\gamma_{\textrm{c}}$ are the corresponding spectral index and cut-off Lorentz factor, respectively. This behavior can be seen in Fig.~\ref{fig:spect_evol}, for instance, in the cases $k_B\overline{ T}/mc^2 \approx 5.2 \times 10^{-2}$ and $3.5 \times 10^{-1}$. For the first temperature we fitted Eq.~\ref{eq:powlaw} using $p \approx 2.2$ and $\gamma_{\textrm{c}} \approx 4.8$ (green dotted line), while for the second temperature we used $p \approx 1.9$ and $\gamma_{\textrm{c}} \approx 35$ (pink dotted line). These fits were obtained using a maximum likelihood analysis considering only particles with energy larger than $10k_B\overline{ T}$.

\subsection{Role of $\omega_{c,0}/\Omega_0$}
\label{sec:rolesepa}

\begin{figure}
    \centering
    \includegraphics[width=0.95\columnwidth]{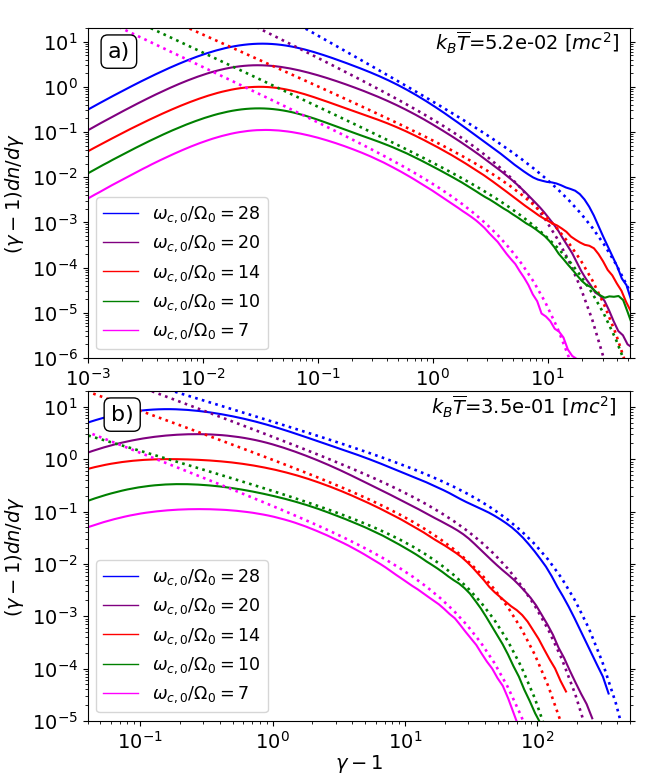}
    \caption{The particle spectra for runs with $\omega_{c,0}/\Omega_0=7, 10, 14, 20$ and 28, and with temperatures $k_B\overline{ T}/mc^2 = 5.2\times 10^{-2}$ (panel $a$) and $3.5\times 10^{-1}$ (panel $b$). For each spectra, we show in dotted lines a power-law fit with an exponential cut-off (as in Eq.~\ref{eq:powlaw}). The normalizations of the spectra are arbitrary.}
%    \vspace{0.1cm}
    \label{fig:spect_mag}
\end{figure}

\begin{figure}
    \centering
    \includegraphics[width=0.95\columnwidth]{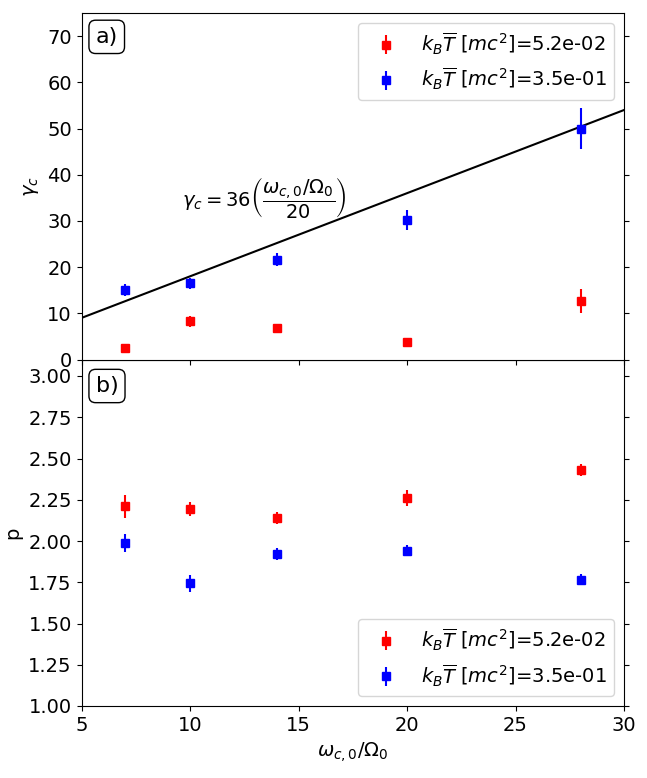}
    \caption{The values of $\gamma_c$ and $p$ obtained from fitting Eq.~\ref{eq:powlaw} to the obtained spectra as a function of $\omega_{c,0}/\Omega_0$. Red and blue squares correspond to $k_B\overline{ T}/mc^2 = 5.2\times 10^{-2}$ and $3.5\times 10^{-1}$, respectively.}
    \label{fig:spect_mag2}
\end{figure}

\noindent The role of the scale-separation ratio $\omega_{c,0}/\Omega_0$ is shown in Fig.~\ref{fig:spect_mag}, which shows the spectra of simulations with $\omega_{c,0}/\Omega_0=7, 10, 14, 20$ and 28, for temperatures $k_B\overline{ T}/mc^2 = 5.2\times 10^{-2}$ (panel $a$) and $3.5\times 10^{-1}$ (panel $b$). For each of these spectra, we show in dotted lines the corresponding fits using power-laws with exponential cut-offs (Eq.~\ref{eq:powlaw}). The dependences of the fitted $\gamma_c$ and $p$ on $\omega_{c,0}/\Omega_0$ are shown in panels $a$ and $b$ of Fig.~\ref{fig:spect_mag2}, respectively. By comparing with the black line in panel $a$ ($\gamma_c = 36 (\omega_{c,0}/\Omega_0)/20$), we see that for the spectra with temperature $k_B\overline{ T}/mc^2 \approx 3.5 \times 10^{-1}$, $\gamma_c$ behaves approximately as $\gamma_{c} \propto \omega_{c,0}/\Omega_0$. For $k_B\overline{ T}/mc^2 \approx 5.2 \times 10^{-2}$, on the other hand, $\gamma_c\sim 4-10$ with no clear dependence on $\omega_{c,0}/\Omega_0$. \newline

\begin{figure}
    \centering
    \includegraphics[width=0.95\columnwidth]{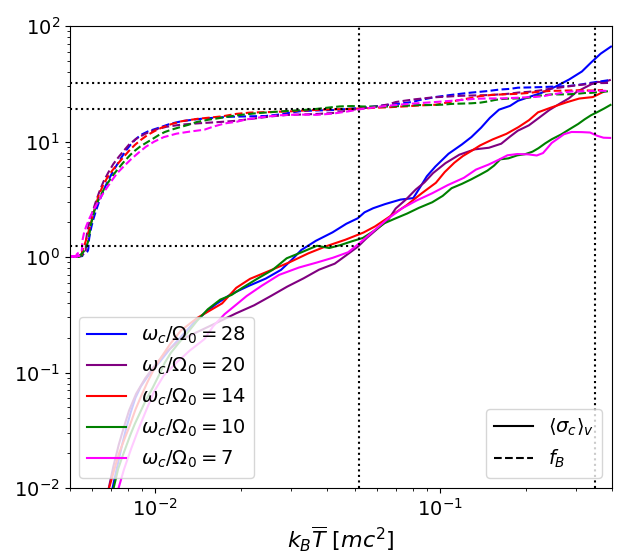}
    \caption{The values of $\langle \sigma_c \rangle_v$ and $ f_B \equiv (\langle B^2\rangle_v)^{1/2}/B_0$ are shown in solid and dashed lines as a function of $\overline{ T }$ and for simulations with $\omega_{c,0}/\Omega_0=7, 10, 14, 20$ and 28.}
    \label{fig:spect_mag3}
\end{figure}

\noindent This discrepancy in how $\gamma_c$ depends on $\omega_{c,0}/\Omega_0$ is likely a manifestation of the underlying acceleration mechanism, which appears to be consistent with the expectation from reconnection driven acceleration. Indeed, the $\gamma_c$ dependence on $\omega_{c,0}/\Omega_0$ for $k_B\overline{ T}/mc^2 \approx 3.5 \times 10^{-1}$ is qualitatively consistente with the pair plasma magnetic reconnection results of \cite{WernerEtAl2016} in the limit of small system size, $L$. These results show power-laws with supra-exponential cut-offs ($dn/d\gamma \propto \gamma^{-p}e^{-\gamma^2/\gamma_{\textrm{c, rec}}^2}$) with $\gamma_{c,\textrm{rec}} \approx 0.1L/\rho_0$, where $\rho_0=mc^2/eB$ and $B$ is the magnitude of the magnetic field in the upstream medium of the reconnecting plasmas. The corresponding value of $L$ in our simulations can be estimated from the power spectrum of the $x-z$ (poloidal) magnetic energy component, $d(|B_x(k)|^2 + |B_z(k)|^2)/d\ln(k)$, for runs with different scale-separation ratios shown in panel $a$ of Fig.~\ref{fig:powerspectrum} (we use the poloidal magnetic field since this is the component that can experience reconnection in 2D). We see that the poloidal spectra peak at $k \sim \Omega_0/2\pi v_{A,0}$ fairly independent of the scale-separation ratio. Thus a reasonable estimate for $L$ is $L \sim 2\pi/k \sim (2\pi)^2v_{A,0}/\Omega_0$. In addition, we can estimate $\rho_0=mc^2/eB\approx c/(\omega_{c,0} f_B)$, where $ f_B$ ($\equiv (\langle B^2\rangle_v)^{1/2}/B_0$) is the root mean square amplification factor of the magnetic field in the disk at a given time. Thus, if reconnection is the main driver of particle acceleration in our runs, $\gamma_c$ should be close to $\gamma_{c,\textrm{rec}}$, which would be given by
\begin{equation}
\gamma_{c,\textrm{rec}}\approx 0.1\frac{L}{\rho_0}\approx 36\Big(\frac{\omega_{c,0}/\Omega_0}{20}\Big)\Big(\frac{ f_B}{30}\Big),
\label{eq:lrho}
\end{equation}
where we have used that $v_{A,0}/c=10^{-2}$ in all our simulations. The value $ f_B $ as a function of $\overline{ T }$ is shown in dashed lines in Fig.~\ref{fig:spect_mag3} for different values of $\omega_{c,0}/\Omega_0$. We see that when $k_B\overline{ T}/mc^2 \approx 3.5 \times 10^{-1}$, $ f_B  \approx 30$, fairly regardless of the scale-separation ratio. This means that the expected $\gamma_{c,\textrm{rec}}$ at $k_B\overline{ T}/mc^2 = 3.5 \times 10^{-1}$ is
\begin{equation}
\gamma_{c,\textrm{rec}}\approx 36\Big(\frac{\omega_{c,0}/\Omega_0}{20}\Big).
\label{eq:gammarec}
\end{equation}

The black line in panel $a$ of Fig.~\ref{fig:spect_mag2} shows the case $\gamma_c = \gamma_{c,\textrm{rec}}$, where $\gamma_{c,\textrm{rec}}$ is given by Eq.~\ref{eq:gammarec}. We see that $\gamma_{c,\textrm{rec}}$ reproduces well the behavior of $\gamma_c$ in our runs with $k_B\overline{ T}/mc^2 = 3.5 \times 10^{-1}$.\newline

\noindent The behavior $\gamma_{c,\textrm{rec}} \approx 0.1L/\rho_0$ expected from reconnection is valid as long as $L/\sigma_c^u\rho_0 \lesssim 40$ \citep{WernerEtAl2016}, where $\sigma_c^u$ corresponds to the cold sigma parameter in the upstream medium of the reconnection simulations. We estimate $\sigma_c^u$ using $\langle \sigma_c\rangle_v$ in our runs when $k_B\overline{T}/mc^2 = 3.5\times 10^{-1}$, which is $\langle \sigma_c\rangle_v \sim 10-50$ for the range of $\omega_{c,0}/\Omega_0$ considered, as shown by the solid lines in Fig.~\ref{fig:spect_mag3}.\footnote{Since we want to estimate the equivalent of the upstream cold sigma parameter $\sigma_c^u$, we are mainly interested in the values of $\sigma_c$ outside the current sheets, where $\sigma_c$ is the largest. Thus, given that  in our runs $\langle \sigma_c\rangle_v$ ($= \langle B^2/4\pi nm^2\rangle_v) >\overline{ \sigma_c}$ ($=\langle B^2\rangle_v/4\pi\langle n\rangle_vm^2$), we are using $\langle \sigma_c\rangle_v$ instead of $\overline{ \sigma_c}$ as our estimate of $\sigma_c^u$.} Thus, using Eq.~\ref{eq:lrho}, we obtain that
\begin{equation}
\frac{L}{\rho_0\sigma_c}\sim 18\Big(\frac{\omega_{c,0}/\Omega_0}{20}\Big)\Big(\frac{f_B}{30}\Big)\Big(\frac{\langle \sigma_c\rangle_v}{20}\Big)^{-1}.
\label{eq:res}
\end{equation}

\noindent Eq.~\ref{eq:res} thus implies that all of our simulations satisfy the restriction $L/\rho_0\sigma_c \lesssim 40$ when $k_B\overline{T}/mc^2 = 3.5\times 10^{-1}$, even in our run with the largest scale-separation ratio, $\omega_{c,0}/\Omega_0=28$. Interestingly, Fig.~\ref{fig:spect_mag3} also shows that, when $k_B\overline{T}/mc^2 = 5.2\times 10^{-2}$, $\langle \sigma_c\rangle_v \sim 1$ and $ f_B  \approx 20$, implying that for that temperature $L/\rho_0\sigma_c \sim 240(\omega_{c,0}/\Omega_0)/20 \gtrsim 40$. This means that, if particle acceleration is driven by magnetic reconnection at $k_B\overline{ T}/mc^2 = 5.2\times 10^{-2}$, $\gamma_c$ should not be proportional to $\omega_{c,0}/\Omega_0$. Instead, a weaker dependence on $L$ is expected, since in that case $\gamma_c$ likely grows more slowly with time as $\gamma_c \propto t^{1/2}$ \citep{PetropoulouEtAl2018,HakobyanEtAl2021}. This possibly explains why we do not observe a clear dependence of $\gamma_c$ on $\omega_{c,0}/\Omega_0$ in the case of $k_B\overline{ T}/mc^2 = 5.2\times 10^{-2}$.\newline

\noindent The values of $p$ for $k_B\overline{ T}/mc^2 = 5.2 \times 10^{-2}$ and $3.5 \times 10^{-1}$ seen in panel $b$ of Fig.~\ref{fig:spect_mag2} are close to $p\sim 2.2$ and $\sim 1.9$, respectively, and do not show a clear dependence on the scale-separation ratio. This is also consistent with acceleration being driven by reconnection. For instance, for $L/\rho_0\sigma_c \gtrsim 40$ and $\sigma_c=3$ (a case close to our results with $k_B\overline{ T}/mc^2 = 5.2 \times 10^{-2}$, where $\langle \sigma_c\rangle_v \sim 1-2$; see Fig.~\ref{fig:spect_mag3}),  \cite{WernerEtAl2016} predicts $p \sim 2.3-2.5$. Whereas for $L/\rho_0\sigma_c \lesssim 40$ and $\sigma_c=10-30$ (close to our results with $k_B\overline{ T}/mc^2 = 3.5 \times 10^{-1}$, where $\langle \sigma_c\rangle_v \sim 10-50$; see Fig.~\ref{fig:spect_mag3}),  the results of \cite{WernerEtAl2016} show $p \sim 1.4-1.9$.

\subsection{Effect of stratification on the acceleration}
\label{sec:impstrat}

\begin{figure}
    \centering
    \includegraphics[width=.95\columnwidth]{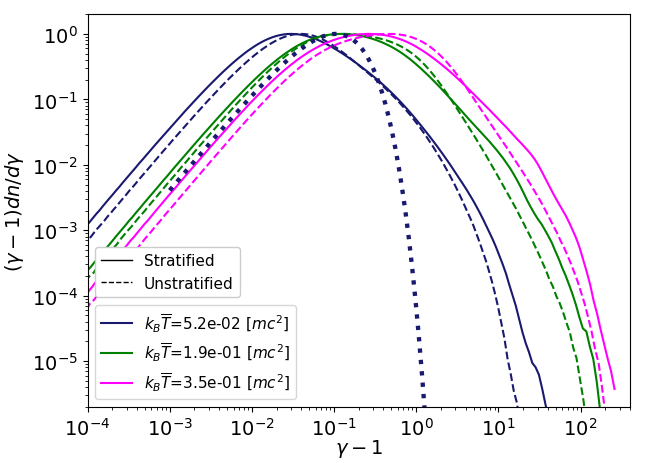}
    \caption{Particle energy distribution for stratified (solid) and unstratified (dashed) runs with the same magnetization $\omega_c/\Omega_0=20$ (runs ST2D-20 and UN2D-20, respectively). Different colors represent different temperatures. The dotted-black line corresponds to a Maxwell-Boltzmann spectrum.}
    \label{fig:spect_evol2}
\end{figure}

\noindent The previous discussion underscores the importance of plasma conditions, in particular $\sigma_c$ and $f_B$, in determining the efficiency of nonthermal particle acceleration. Since these conditions vary significantly between stratified and unstratified simulations (as shown by Fig.~\ref{fig:volav_su}), we expect the acceleration efficiency in these two types of runs to be different. Fig \ref{fig:spect_evol2} compares spectra from run ST2D-20 with the equivalent spectra in the unstratified run UN2D-20 at the same values of $\overline{ T}$. We see that the spectra in the unstratified run are always softer than in the stratified run ST2D-20. This is consistent with the fact that, for a given temperature, in run UN2D-20 the value of $\overline{\sigma}_c$ is smaller than in the run ST2D-20, which favors harder nonthermal acceleration (as seen in panel $c$ of Fig.~\ref{fig:volav_su})

\subsection{Acceleration in 2D vs 3D}
\label{sec:2d3d}

\begin{figure}
    \centering
    \includegraphics[width=0.945\columnwidth]{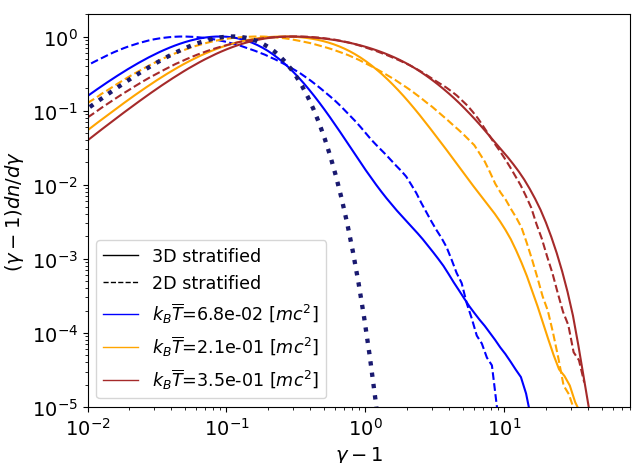}
    \caption{Spectra from simulations in 2D (dashed) and 3D (solid), both with a scale-separation ratio $\omega_{c,0}/\Omega_0=3.5$ (runs ST2D-3.5 and ST3D-3.5., respectively). The different colors represent  $k_B\langle T\rangle/mc^2 = 6.8 \times 10^{-2}$, $2.1 \times 10^{-1}$ and $3.5 \times 10^{-1}$.}
    \label{fig:spectrum2d3d}
\end{figure}

\noindent In Fig.~\ref{fig:spectrum2d3d} we compare spectra from the 2D and 3D simulations ST2D-3.5 and ST3D-3.5, both with a scale-separation ratio $\omega_{c,0}/\Omega_0=3.5$, for  $k_B\overline{T}/mc^2 = 6.8 \times 10^{-2}$, $2.1 \times 10^{-1}$ and $3.5 \times 10^{-1}$. As expected from our previous discussion on the dependence of $\gamma_c$ on $\omega_{c,0}/\Omega_0$, the 2D run ST2D-3.5 should produce a nonthermal tail of rather short extension, which is what we see in Fig.~\ref{fig:spectrum2d3d}. However, it is still interesting to verify whether its main features are reproduced in the 3D run ST3D-3.5. We see that, although the spectra show somewhat different shapes, they both feature nonthermal tails with similar maximum energies. In particular, when $k_B\overline{ T}/mc^2 = 3.5 \times 10^{-1}$, the 2D and 3D spectra look very similar, suggesting that 3D effects maintain the main particle accelerating properties of the MRI turbulence in the stratified setup.\newline
\noindent Even though the nonthermal particle behavior in our runs suggests a significant role of magnetic reconnection in the acceleration of particles, our simulations may be subject to effects that are not present in previous magnetic reconnection studies. These include particle escape from the disk, stochastic acceleration by the MRI turbulence \citep[e.g.,][]{KimuraEtAl2019,SunEtAl2021}, and the action of various kinetic instabilities that may contribute to field dissipation and/or particle acceleration, including, e.g., the drift kink instability \citep{ZenitaniEtAl2007} and the ion-cyclotron instability \citep{LeyEtAl2019}. We thus defer to future research a detailed determination of the dominant acceleration process(es) as well as the role of the scale-separation ratio by including 2D and 3D runs with larger values of $\omega_{c,0}/\Omega_0$.

\section{Conclusions}
\label{sec:conclu}
\noindent In this work we have studied the effect of stratification on the collisionless MRI using 2D and 3D PIC simulations. Comparing 2D stratified and unstratified runs, we found that stratification affects the evolution of the disk conditions, due to the presence of outflows and disk expansion, leading to a decrease in the amplification of magnetic field energy density in the turbulent non-linear MRI regime. However, the expansion of the disk also decreases the plasma pressure and density, resulting in a highly magnetized disk, with smaller $\beta$ and larger cold magnetization parameter $\sigma_c$ compared to the unstratified case. Indeed, in the nonlinear regime the disk is magnetic-pressure supported with $\beta \sim 0.4$, which is a factor $\sim 5$ smaller than the value reached by its unstratified counterpart. Our stratified simulations do not exhibit a discernible low beta corona separated from the disk. 
%This is likely caused by the relatively small initial beta ($\beta_0=100$) in our disk, as suggested by previous 3D MHD simulations of stratified disks with similar $\beta_0$ \citep[e.g.,][]{SalvesenEtAl2016}. 
In the disk region, our runs also give rise to a significant large scale and predominantly toroidal dynamo-like field $\mathbf{B}^D$ ($\equiv\langle \mathbf{B} \rangle_x$ in 2D), whose dominant scale length follows the disk scale height. Although a large scale $\langle \mathbf{B} \rangle_x$ field also appears in the 2D unstratified case, its scale length is $\sim 4$ times smaller. The increased magnetization of our 2D stratified runs produces an effective viscosity $\alpha$ in the disk that reaches $\alpha \sim 1$, which is a factor $\sim 5$ larger than in the equivalent unstratified case. This viscosity is dominated by the Maxwell stress, $\alpha_M$, with a small contribution of the anisotropic stress, $\alpha_A$. This small $\alpha_A$ is consistent with the regulation of pressure anisotropies by kinetic microinstabilities in the low $\beta$ regime.\newline 

\noindent Even though our 2D and 3D stratified simulations produce similar results, some differences are present. In order to assess them, we compared 2D and 3D runs focusing on a specific case with small scale separation, $\omega_{c,0}/\Omega_0=3.5$. In the early phase of the non-linear MRI stage (i.e., $\sim 1-2$ orbits after the triggering of the instability), 3D simulations exhibit a significantly lower amplification of the magnetic field energy density compared to their 2D counterpart, consistently with a more  efficient reconnection of the toroidal magnetic field component in 3D \citep[as shown in the recent work of][]{BacchiniEtAl2022}. This primarily affects the effective viscosity $\alpha$ and the plasma $\beta$, which are, respectively, $\sim 3-4$ times smaller and larger in the 3D case. However, after this initial stage (at $t \sim 4$ [$2\pi/\Omega_0$]), our 2D and 3D simulations are more similar, with $\beta$ reaching essentially the same values and the effective viscosity $\alpha$ being only $\sim 2$ times smaller in 3D (in 3D, $\alpha \approx 0.5$ during the whole nonlinear MRI stage). This transition at $t \sim 4$ [$2\pi/\Omega_0$] occurs because of the growing importance of the large scale dynamo-like field $\mathbf{B}^D$ ($\equiv\langle \mathbf{B} \rangle_{x-y}$ in 3D). Indeed, after an initial stage in which the turbulent field $\mathbf{B}^T$ ($=\mathbf{B}-\mathbf{B}^D$) dominates, the dynamo-like field becomes larger than the turbulent field in the 3D runs while in 2D it reaches values comparable to the turbulent field. Since the dynamo field has almost the same amplitude in 2D and 3D, the total fields in these two types of runs differ by a small amount after $t \sim 4$ [$2\pi/\Omega_0$]. Also, in this dynamo-dominated period, the 3D viscosity is mainly produced by the dynamo-like field, while in 2D the turbulent and dynamo fields contribute comparably to $\alpha$. In 3D the disk viscosity is also dominated by the Maxwell stress, $\alpha_M$, with a small contribution from the anisotropic stress, $\alpha_A$. This is also consistent with the action of pressure anisotropy-driven kinetic microinstabilities in the 3D case, as it occurs in 2D. Our 2D and 3D results in terms of $\alpha$, $\beta$ and dynamo-like field behaviors are reasonably consistent with previous 3D MHD simulations of stratified disks with similar initial conditions \citep[e.g.,][]{BaiEtAl2013,SalvesenEtAl2016}. \newline

\noindent In terms of particle acceleration, in our 2D runs we find that the particle spectra in the nonlinear MRI stage follow power-laws with exponential cut-offs, with power-law indices $p\approx 2.2-1.9$ for disk temperatures $\sim 0.05-0.3\, mc^2$. Additionally, depending on the value of $\sigma_c$ during the nonlinear MRI stage, the maximum energy attained by the particles is either proportional to the scale separation $\omega_{c,0}/\Omega_0$ or fairly independent of this parameter, which appears to be consistent with previous magnetic reconnection studies \citep{WernerEtAl2016}. Particle acceleration in our 2D unstratified runs appears to be less efficient than in the analogous stratified case. This is likely due to the smaller cold magnetization parameter $\sigma_c$ attained in the unstratified simulations. Furthermore, the particle acceleration observed in our 2D run with $\omega_{c,0}/\Omega_0=3.5$ is well reproduced by its analogous 3D simulation, suggesting that 3D effects should maintain most of the acceleration properties of the MRI turbulence. However, 3D runs with larger scale-separation ratio are needed to confirm this trend.\newline

\noindent In summary, our results suggest that including disk stratification in shearing-box PIC simulations of the MRI is important for studying its saturation, effective viscosity generation and particle acceleration physics. Interestingly, 2D and 3D simulations give quite similar results for the scale-separation ratios used in this work, especially after the magnetic field energy becomes dominated by a large-scale, dynamo-like field (which occurs $\sim 1-2$ orbits after the triggering of the instability). We leave for future work the clarification of the effect of larger scale-separation ratios in 3D, as well as disentangling the underlying mechanism(s) for particle acceleration. We also note that our results refer to a specific case of initial plasma conditions. Further research is thus needed to clarify the effects of changing the initial $\beta$ and/or temperature in the disk, potentially leading to a more distinct differentiation between an unmagnetized disk and a magnetized corona \citep[e.g.,][]{SalvesenEtAl2016}. Investigating the effect of more realistic mass ratios on the dynamic and thermodynamic properties of the collisionless MRI is also deferred to future work.\newline

\section*{Acknowledgements}

\noindent A. Sandoval acknowledges support from the Center for Excellence in Astrophysics and Associated Technologies (CATA) through ANID, BASAL, FB210003. M. Riquelme thanks support from a Fondecyt Regular Grant No. 1191673 and from CONICYT/Quimal 190011. A. Spitkovsky acknowledges the support of NSF grants PHY-2206607 and AST-1814708. F. Bacchini acknowledges support from the FED-tWIN programme (profile Prf-2020-004, project ``ENERGY'') issued by BELSPO. The computational resources and services used in this work were provided by the National Laboratory for High Performance Computing (NLHPC) of the Center for Mathematical Modeling of University of Chile (ECM-02) and the VSC (Flemish Supercomputer Center), funded by the Research Foundation -- Flanders (FWO) and the Flemish Government -- department EWI. 

\section*{Data Availability}
%%%%%%%%%%%%%%%%%%%%%%%%%%%%%%%%%%%%%%%%%%%%%%%%%%

\noindent The data underlying this article will be shared on reasonable request
to the corresponding author.



\bibliographystyle{mnras}

%%%%%%%%%%%%%%%%%%%%%%%%%%%%%%%%%%%%%%%%%%%%%%%%%%

%%%%%%%%%%%%%%%%% APPENDICES %%%%%%%%%%%%%%%%%%%%%

%/appendix



%%%%%%%%%%%%%%%%%%%%%%%%%%%%%%%%%%%%%%%%%%%%%%%%%%

% Don't change these lines
\bsp	% typesetting comment
\label{lastpage}
\end{document}